\documentclass[aoas,nameyear,MSNbibl,dvips]{arximspdf}
\usepackage{dcolumn,multirow}
\usepackage{graphicx}

\doi{10.1214/11-AOAS454}
\volume{5}
\issue{3}
\pubyear{2011}
\firstpage{1948}
\lastpage{1977}

\makeatletter
\newcommand{\ctn}{\citet}
\newcommand{\gfrac}[2]{#1/#2}

\newcolumntype{d}[1]{D{.}{.}{#1}}

\newcommand{\bTheta}{\bolds{\Theta}}
\newcommand{\bLambda}{\bolds{\Lambda}}
\newcommand{\bSigma}{\bolds{\Sigma}}
\newcommand{\bmu}{\bolds{\mu}}
\newcommand{\bC}{\mathbf{C}}
\newcommand{\bG}{\mathbf{G}}
\newcommand{\bS}{\mathbf{S}}
\newcommand{\bY}{\mathbf{Y}}
\newcommand{\bZ}{\mathbf{Z}}

\let\epsilon\varepsilon

\let\sv@tabnotetext\tabnotetext
  \let\sv@tabnotemark@fmt\tabnotemark@fmt
   \long\def\legend#1{{\let\tabnote@indent\leavevmode\sv@tabnotetext[]{}{#1}}}

\makeatother

\begin{document}
\begin{frontmatter}

\title{On Bayesian ``central clustering'': Application to landscape
    classification of Western Ghats}
\runtitle{Bayesian central clustering}

\begin{aug}
\author[A]{\fnms{Sabyasachi} \snm{Mukhopadhyay}\ead[label=e1]{sabstat123@gmail.com}},
\author[A]{\fnms{Sourabh} \snm{Bhattacharya}\corref{}\ead[label=e2]{bhsourabh@gmail.com}}\\
and
\author[B]{\fnms{Kajal} \snm{Dihidar}\ead[label=e3]{dkajal@isical.ac.in}}
%
%
\runauthor{S. Mukhopadhyay, S. Bhattacharya and K. Dihidar}
\affiliation{Indian Statistical Institute}
\address[A]{S. Mukhopadhyay\\
S. Bhattacharya\\
Bayesian and Interdisciplinary Research Unit\\
Indian Statistical Institute\\
203 B. T. Road\\
Kolkata 700108\\
India\\
\printead{e1}\\
\hphantom{E-mail: }\printead*{e2}} 
\address[B]{K. Dihidar\\
Applied Statistics Unit\\
Indian Statistical Institute\\
203 B. T. Road\\
Kolkata 700108\\
India\\
\printead{e3}}
\end{aug}

\received{\smonth{5} \syear{2010}}
\revised{\smonth{12} \syear{2010}}

%
\begin{abstract}
Landscape classification of the well-known biodiversity hotspot,
Western Ghats (mountains), on the west coast of India,
is an important part of a world-wide program of monitoring
biodiversity. To this end, a massive vegetation data set,
consisting of 51,834 4-variate observations has been clustered into
different landscapes by Nagendra and Gadgil
[\textit{Current Sci.} \textbf{75} (\citeyear{Nagendra98}) 264--271].
But a study of such importance may be affected by nonuniqueness of
cluster analysis and the lack
of methods for quantifying uncertainty of the clusterings obtained.

Motivated by this applied problem of much scientific importance, we
propose a new methodology for obtaining the
global, as well as the local modes
of the posterior distribution of clustering, along with the desired
credible and ``highest posterior density'' regions in a nonparametric
Bayesian framework.
To meet the need of an appropriate metric for computing the distance
between any two clusterings,
we adopt and provide a much simpler,
but accurate modification of the metric proposed in [In
\textit{Felicitation Volume in Honour of Prof. B. K. Kale} (\citeyear{Ghosh08}) MacMillan].
A very fast and
efficient Bayesian
methodology, based on [\textit{Sankhy\=a Ser. B} \textbf{70}
(\citeyear{Bhatta08}) 133--155], has been utilized to
solve the computational
problems associated with the massive data and
to obtain samples from the posterior distribution of clustering on
which our proposed methods of
summarization are illustrated.

Clustering of the Western Ghats data using our methods yielded
landscape types different from
those obtained previously, and provided interesting insights concerning
the differences
between the results obtained by Nagendra and Gadgil
[\textit{Current Sci.} \textbf{75} (\citeyear{Nagendra98}) 264--271] and us.
Statistical implications of the differences
are also discussed in detail, providing interesting insights into
methodological concerns of the
traditional clustering methods.
\end{abstract}

%

\end{frontmatter}

\section{Introduction}
\label{sec:intro}

Nagendra and Gadgil (\citeyear{Nagendra98}) (henceforth, NG) consider a broad scale mapping of
the Western Ghats of India,
one of the biodiversity hotspots of the world, into different
landscape types based on satellite imagery. This exercise is a part of a
much bigger program
related to monitoring and
assessment of measures of conservation.
Remote sensing-based identification of landscapes of
different types in important biodiversities such as the Western Ghats
is necessary for constituting a basis for organized programs of field
samplings (see NG, page~270, for the detailed procedure of field
sampling), and to create
administrative divisions such as taluks and
districts and bioclimatic zones. Formation of administrative divisions,
unlike bioclimatic zones,
need not be directly based on natural variation,
but these reflect natural topographic and climatic variation to some extent.
Using a massive vegetation data set based on satellite images, which
consists of 51,834 4-variate
observations, NG obtained a clustering of the data using a
deterministic algorithm very similar to the $K$-means
algorithm [see, e.g., \ctn{Hartigan75}], and related the
different clusters to landscape types
of varying attributes.\looseness=-1

However, the existing clustering algorithms, including that used by NG,
have some serious disadvantages,
which we outline in Section \ref{subsec:disadvantages}.
These are likely to severely affect the scientific results of important
studies, such as that undertaken by NG.
This motivated us to propose new methods of clustering; the results we
obtained with
our methods, apart from some broad similarities, differed
nonnegligibly in details from those obtained by NG,
vindicating our purpose and efforts of methodological
development.

\subsection{Disadvantages of existing clustering methods and the need
for new methods}\label{subsec:disadvantages}
By clustering we mean partitioning the observed data into several
different classes or clusters.
Although the statistical community is very much aware of the
definition, clustering of a particular data set
is usually taken to mean a particular, perhaps unique, partitioning of
the data into various clusters, the number
of clusters being known, or at least determined using statistical
techniques or information based on scientific knowledge.

\subsubsection{Disadvantages of deterministic clustering algorithms}\label{subsubsec:disadvantages_deterministic}

But well established clustering algorithms, such as the $K$-means algorithm,
may yield different clusterings under
different starting points. This leads to nonunique clusterings of the
same data set, which, in turn,
begs the question of ascertaining the uncertainties of the clusterings
obtained. However, deterministic
(nonprobabilistic) clustering
algorithms provide no means of quantification of such uncertainty.
Moreover, in these
algorithms one must somehow fix the number of clusters, and the basis of
such fixing is often not clear cut.

\subsubsection{Disadvantages of classical model-based clustering}
\label{subsubsec:disadvantages_classical}
Probabilistic\break model-based clustering methods within the classical
framework
provide an estimate of the data clustering, along with the parameter\vadjust{\eject}
estimates, by maximizing
the likelihood [see, e.g., \ctn{Fraley99} and \ctn{Fraley02}].
As before, the number of clusters
is assumed known, and uncertainties
about clustering estimation and the number of clusters are not taken
into account even in this approach.

\subsubsection{Disadvantages of Bayesian clustering}
\label{subsubsec:disadvantages_bayesian}
In contrast to the deterministic and classical model-based clustering
methods, the Bayesian paradigm offers attractive ways
to assign probabilities to plausible clusterings,
while allowing even for the number of clusters to be a random variable,
using the Dirichlet process mixture
[see, e.g., \ctn{Ferguson73} and \ctn{Antoniak74}] approach
of \ctn{Escobar95} (henceforth, EW) and the reversible jump Markov
chain Monte Carlo approach (RJMCMC)
of \ctn{Richardson97}.
But in spite
of the promise held out by the Bayesian paradigm and these pioneering
approaches, summarization and addressing the posterior uncertainty
of clusterings seem to be somewhat neglected so far. The maximum a
posteriori (MAP) estimate of clustering,
often available for Bayesian mixture models [see, e.g., \ctn
{Dahl09} and the references therein],
is not supplemented with appropriate quantification of uncertainty.
A further disadvantage of the aforementioned Bayesian methods is their
inability to handle massive data sets.
Indeed, implementation of these methods turned out to be infeasible in
the case of the massive, multivariate,
Western Ghats data.

\subsection{Overview of the new contributions of this paper}
\label{subsec:new_contributions}

\subsubsection{Methodological contributions}
\label{subsubsec:methodological_contributions}
In this paper we attempt to address the important issue
of summarizing and quantifying uncertainty of posterior distributions
of clusterings. In particular, we propose a novel
approach to determination of the global mode, as well as the local
modes, of the posterior of clusterings in a Bayesian
nonparametric setup, based on a Dirichlet process prior. We refer to
such modes, thought of as summaries or representatives of the
posterior, as ``central clusterings.'' Much more importantly,
we show that, using our approach to obtaining central clusterings, any
desired credible or highest posterior density (HPD) regions
[see, e.g., \ctn{Berger85}] are also available, completely
quantifying uncertainty of the posterior of clusterings.
Necessary for these developments is an appropriate metric to compute
the distance between any two clusterings.
We adopt the metric proposed in \ctn{Ghosh08}, but since it is
computationally expensive, we
propose a simple, albeit
accurate, approximation to the metric, which we use to compute
summaries of the posterior distribution of clusterings.
We illustrate our proposed methods with simulated data,
and also apply these to the
Western Ghats data set. Although implementation of the established
Bayesian methods are rendered infeasible by the massiveness
of the data,
we solve this massive data analysis problem by employing a~very fast
and efficient Bayesian methodology,
first proposed in \ctn{Bhatta08} (henceforth, SB).

\subsubsection{Overview of statistical and ecological insights gained
by analyzing the Western Ghats data}
\label{subsubsec:data_insights}
The results of our application to the Western Ghats data revealed two
modal clusterings, in contrast with the
single clustering obtained by NG. Moreover, the $K$-means clustering,
which can
be thought of as a proxy to that obtained by NG, does not fall within
our 95\% HPD or credible regions, raising doubts
about the validity of NG's adopted methodology and the results, even
though their number of clusters matched ours with the highest
posterior probability. However, the $K$-means clustering fell within
our 95\% HPD and credible regions when these are
constructed conditional on the same number of clusters as the $K$-means
clustering. These, which are discussed in detail
in the paper, are consequences of the failure of the deterministic
algorithm to take account of uncertainty
in the number of clusters. Detailed comparisons between the clusterings
we obtained by our methods and the $K$-means clustering are made
statistically, as well as with respect to the
landscape types associated with each cluster of each clustering. In
fact, the attributes of the
landscape types obtained by our methods
turned out to be generally different from those obtained by~NG.

The rest of our paper has been arranged as follows.
In Section \ref{sec:dp} we describe the Bayesian model based on SB
used to analyze the Western Ghats
data. Methods for summarizing general posterior distributions of clustering
are introduced in Section~\ref{sec:central_clustering}. In Section
\ref{sec:metric} we
provide an overview of the clustering metric of \ctn{Ghosh08}, propose
an accurate and computationally
simple approximation to the clustering metric, and study its properties.
Applications of our methods to the Western Ghats data are illustrated
in Section \ref{sec:real_data}.
Detailed interpretation of the clustering results in terms of different
landscape types are presented
in Section \ref{sec:interpretation}.
Finally, we conclude in Section \ref{sec:conclusion}.
Additional derivations and further details on
experiments and data analyses are provided in
the supplement \ctn{Mukh11}, whose sections, figures and tables have
the prefix
``S-'' when referred to in this paper.

\section{Mixture model of SB}
\label{sec:dp}

Following SB, we model the $d (\geq1)$-variate observation
$\mathbf{y}_i$ of the complete data set $\bY= \{\mathbf{y}_1,\ldots,\mathbf{y}_n \}$
as a mixture of normals with maximum number of components $M$, as follows:
%
\begin{equation}
[\mathbf{y}_i\mid\bTheta_M]=\frac{1}{M}\sum_{j=1}^M\frac
{|\bLambda_j|^{\gfrac{1}{2}}}{(2\pi)^{\gfrac{d}{2}}}\exp \biggl\{
-\frac{1}{2} (\mathbf{y}_i-\bmu_j )'\bLambda_j
(\mathbf{y}_i-\bmu_j ) \biggr\}.
\label{eq:mult_normixture}
\end{equation}
Here $\bTheta_M=\{\bolds{\theta}_1,\ldots,\bolds{\theta
}_M\}$, with $\bolds{\theta}_j=(\bmu_j,\bLambda_j)$, where
$\bLambda_j=\bSigma^{-1}_j$,
are samples drawn from a Dirichlet process [see, e.g., \ctn
{Ferguson73}, EW]:
\begin{eqnarray}
\bolds{\theta}_j&\stackrel{\mathrm{i.i.d.}}{\sim}&\bG,\nonumber\\
\bG&\sim& \operatorname{DP}(\alpha\bG_0).\nonumber
\end{eqnarray}
We assume that under $\bG_0$,
%
\begin{eqnarray}
 [\bLambda_j ]&\sim&\operatorname{Wishart}_d \biggl(\frac{s}{2},\frac
{\bS}{2} \biggr),\label{eq:wishart_d}\\
 {}[\bmu_j\mid\bLambda_j ]&\sim& N_d (\bmu_0,\psi
\bLambda^{-1}_j ).\label{eq:normal_d}
\end{eqnarray}
Due to the discreteness
of the prior distribution $\bG$,
the parameters $\bolds{\theta}_{\ell}$ are coincident with
positive probability.
This discreteness property of Dirichlet processes ensures that (\ref
{eq:mult_normixture}) reduces to the form
%
\begin{equation}
[\mathbf{y}_i\mid\bTheta_M]=\sum_{j=1}^p\pi_j\frac{|\bLambda
^*_j|^{\gfrac{1}{2}}}{(2\pi)^{\gfrac{d}{2}}}\exp \biggl\{-\frac
{1}{2} (\mathbf{y}_i-\bmu^*_j )'\bLambda^*_j
(\mathbf{y}_i-\bmu^*_j ) \biggr\},
\label{eq:mult_normixture2}
\end{equation}
where $ \{\bolds{\theta}^*_1,\ldots,\bolds{\theta
}^*_p \}$ are $p$ distinct components of $\bTheta_M$ with
$\bolds{\theta}^*_j$ occuring $M_j$ times,
and $\pi_j=M_j/M$.
Thus, our model is also variable dimensional but avoids complexities as
in RJMCMC.

Introducing the allocation variables $\bZ=(z_1,\ldots,z_n)'$,
we can represent (\ref{eq:mult_normixture})
as follows:

For $i=1,\ldots,n$ and $j=1,\ldots,M$,
%
\begin{eqnarray}
 [\mathbf{y}_i\mid z_i=j,\bTheta_M ]&=&\frac
{|\bLambda_j|^{\gfrac{1}{2}}}{(2\pi)^{\gfrac{d}{2}}}\exp \biggl\{
-\frac{1}{2} (\mathbf{y}_i-\bmu_j )'\bLambda_j
(\mathbf{y}_i-\bmu_j )\biggr\},
\label{eq:y_given_z}\\
{}[z_i=j ]&=&\frac{1}{M}.
\label{eq:latent_z}
\end{eqnarray}
We note here that the number of components is not the same as the
number of clusters in the case of SB's model, although
the maximum number of components and the maximum number of clusters are
the same. This is because
there may be empty components, to which no data may be allocated. This
is unlike the case of EW's model, where empty components
can not exist, so that the number of components is the same as the
number of clusters. This can be seen
by letting $M=n$ and $z_i=i$ for $i=1,\ldots,n$ in SB's model, which
then reduces to EW's model, where $z_i=i$
rules out the existence of empty components. In SB's model we say that
the $i$th data point belongs to the
$j$th cluster
if $\bolds{\theta}_{z_i}=\bolds{\theta}^*_j$, where
$\bolds{\theta}^*_j$ is the $j$th distinct component in $\bTheta_M$.

Letting $\{\bolds{\theta}^*_1,\ldots,\bolds{\theta}^*_k\}
$ denote the distinct components in $\bTheta_M$, let us define
the configuration vector $\bC=\{c_1,\ldots,c_M\}$, where, for
$j=1,\ldots, M$ and $\ell=1,\ldots,k$, $c_j=\ell$
if and only if $\bolds{\theta}_j=\bolds{\theta}^*_{\ell
}$. With this setup, given the hyperparameters
$\bmu_0$ and $\psi$, two versions
of Gibbs sampling are possible: one version
updates $(\bZ,\bC,k,\bolds{\theta}^*_1,\ldots,\bolds{\theta}^*_k,\alpha)$ in succession, while another marginalizes the model
with respect to
$\{\bolds{\theta}^*_1,\ldots,\bolds{\theta}^*_k\}$ and
updates in succession the reduced set of random variables
$(\bZ,\bC,k,\alpha)$.
These two versions of Gibbs sampling are provided in Sections~S-1 and
S-2, respectively.
Details on the priors are provided in Section \ref{subsec:prior}.

It is to be noted that the $K$-means
algorithm of NG is a special case of SB's nonparametric Bayesian model.
It corresponds to taking $M=n$, $z_i=i$ for $i=1,\ldots,n$,
$\bSigma_j=\sigma^2\mathbf{I}$; $j=1,\ldots,M(=n)$ in the
above-described model, where $\mathbf{I}$ is the identity matrix
and $\sigma^2$ is assumed to be known; moreover, it assumes $\bG_0$,
the base measure
for $\bmu_j$ to be
noninformative and that the Bayesian model is conditioned on $k$
clusters, where $k$ is assumed
to be known.
See Section S-3 for a proof of the result.

\section{Summarization of the posterior distribution of clustering}
\label{sec:central_clustering}

It is evident from the previous section that the clustering and the
number of clusters vary in each iteration of the Gibbs sampling algorithm.
In fact, even if the number of clusters are the same in any two
iterations, the corresponding clusterings are still different. The
statistician is faced with
the question of obtaining a summary of all the clusterings obtained
from the Gibbs sampling algorithm, since a representative of all the clusterings
produced (the posterior distribution of clustering) is usually of
scientific interest. Observe that this problem is much more difficult
as compared to summarization
of posterior distribution of a parameter. In the case of a parameter,
the posterior distribution (even sampling-based) can be summarized by its
posterior mean or mode (analytical or sample-based). Similarly, desired
credible regions are readily available. On the other hand, it is just
not possible
to take means of clusterings produced; the mean will give rise to an
$M$-component clustering,
even if all the clusterings consist of less than $M$ clusters.
Moreover, the clusterings are permutation-invariant, a fact that simple
means or modes fail
to take account of. Construction of credible regions of such an
abstract concept poses even more difficulties. We propose a methodology
to usefully interpret posterior distributions of clusterings.
For this we need to introduce the concept of ``central clustering,''
which we do below.

\subsection{Definition of central clustering}
\label{subsec:exact_central_clustering}

Motivated by the definition of mode in the case of parametric
distributions, we define that clustering
$C^*$ as ``central,'' for a given small $\epsilon>0$, satisfies the
following equation:
%
\begin{equation}
P \bigl( \{C\dvtx d(C^*,C)<\epsilon \} \bigr)=\sup_{C'} P
\bigl( \{C\dvtx d(C',C)<\epsilon \} \bigr).
\label{eq:central_cluster}
\end{equation}
Note that $C^*$ is the global mode of the posterior distribution of
clustering as $\epsilon\rightarrow0$.
Thus, for a sufficiently small $\epsilon>0$, the probability of
an\vadjust{\eject}
$\epsilon$-neighborhood of an
arbitrary clustering $C'$, of the form $ \{C\dvtx d(C',C)<\epsilon
 \}$,
is highest when $C'=C^*$, the central clustering.

The above definition will hold for all positive $\epsilon$ if the
distribution of
clustering is unimodal.
However, for multimodal distributions of clustering, the central
clustering will not remain the same for all
such $\epsilon$. For instance, due to discreteness of the distribution
of clusterings,
for some $\epsilon$, the neighborhood of the global mode may contain
just a
few clusterings (other than the global mode), while for the same
$\epsilon,$ the neighborhood of some local mode may
contain many more clusterings. This would yield the local mode as
another central clustering.
Thus, by allowing $\epsilon$ to vary uniformly over $(0,1)$, all the
modes of the posterior
of clustering can be detected, including the global mode, the latter
obtained by letting $\epsilon\rightarrow0$.

In (\ref{eq:central_cluster}), $d$ is a suitably chosen metric that is
capable of measuring distances
between any two clusterings, appropriately taking account of the
different number of clusters in each clustering
and invariance of a clustering with respect to permutation of its
components. We note that, with two different sets of
mean parameter vectors, $ \{\bmu^{(k)}_1,\ldots,\bmu
^{(k)}_n \}$ and $ \{\bmu^{(\ell)}_1,\ldots,\bmu^{(\ell
)}_n \}$,
the simple Euclidean
distance between two corresponding clusterings $C^{(k)}$ and~$C^{(\ell)}$, defined by
%
\begin{equation}
d\bigl(C^{(k)},C^{(\ell)}\bigr)
=\sqrt{\sum_{i=1}^n\sum_{j=1}^p  \bigl(\mu^{(k)}_{ij}-\mu^{(\ell
)}_{ij} \bigr)^2},
\label{eq:euclidean_metric}
\end{equation}
is an easily computable option, but it does not take account of the features
discussed. It is thus important to introduce a more specialized metric
that is capable of addressing the
problems, and yet remains computationally inexpensive. We discuss one
such choice, illustrated in detail in Section~\ref{sec:metric}.

It is important to observe that, even with a suitable metric $d$ and
any choice of $\epsilon$, it is not
possible to obtain the central clustering $C^*$ without resorting to
empirical methods.
Indeed, it is not possible to evaluate either side of (\ref
{eq:central_cluster}) analytically.
We thus consider an alternative, empirical definition conditional upon
availability of MCMC
samples of clusterings $ \{C^{(1)},C^{(2)},\ldots,C^{(N)} \}
$, following which one can
determine an approximate central clustering $C^*$.

\subsection{Empirical definition of central clustering}
\label{subsec:empirical_central_clustering}

We define that clustering $C^{(j)}$ as ``approximately central,'' for a
given small $\epsilon>0$, satisfies
the following equation:
%
\begin{equation}
C^{(j)}=\arg\max_{1\leq i\leq N}\frac{1}{N}\#\bigl \{C^{(k)};1\leq
k\leq N\dvtx d\bigl(C^{(i)},C^{(k)}\bigr)<\epsilon \bigr\}.
\label{eq:empirical_central_cluster}
\end{equation}
The central clustering $C^{(j)}$ is easily computable, given $\epsilon
>0$ and a suitable metric~$d$. Also,
by the ergodic theorem, as $N\rightarrow\infty$ the empirical central
clustering $C^{(j)}$ converges almost surely to the exact
central clustering $C^*$.

\subsection{Construction of desired credible regions of clusterings}
\label{subsec:credible}
Given a central clustering $C^{(j)}$, one can then obtain, say, an approximate
95\% posterior density credible region
as the set $ \{C^{(k)};1\leq k\leq N:d(C^{(k)},C^{(j)})<\epsilon
^* \}$, where $\epsilon^*$
is such that
%
\begin{equation}
\frac{1}{N}\# \bigl\{C^{(k)};1\leq k\leq N\dvtx d\bigl(C^{(k)},C^{(j)}\bigr)<\epsilon
^* \bigr\}\approx0.95.
\label{eq:cred}
\end{equation}
In (\ref{eq:cred}) $\epsilon^*$ can be chosen adaptively by starting
with $\epsilon^*=0$ and then slightly
increasing $\epsilon^*$ by a quantity $\zeta$ until (\ref{eq:cred})
is satisfied. For our Western Ghats example
we chose $\zeta=10^{-10}$.
Approximate highest posterior density (HPD) regions
can be constructed by taking the union of the highest density regions.
We next discuss an adaptive methodology
for constructing HPD regions.

\subsection{Construction of desired HPD regions of clusterings}
\label{subsec:hpd}
Assume that there are $k$ modes, $\{C^*_1,\ldots, C^*_k\}$, obtained
by varying $\epsilon$ of
the neighborhoods $\{C\dvtx  d(C\dvtx C^{(i)})<\epsilon\};i=1,\ldots,N$,
uniformly over the interval $(0,1)$,
and following the principle described in Section \ref
{subsec:exact_central_clustering}.
Also consider $k$ $\epsilon^*$'s, $\{\epsilon^*_1,\ldots,\epsilon
^*_k\}$.
Consider the regions $S_j=\{C\dvtx d(C^*_j,C)<\epsilon^*_j\}$;$ j=1,\ldots
,k$. Set, initially, $\epsilon^*_1=\epsilon^*_2=\cdots=\epsilon^*_k=0$.
\begin{longlist}[(iii)]
\item[(i)] For $i=1,\ldots,N$, if the $i$th MCMC realization
$C^{(i)}$ does not fall in~$S_j$ for some $j$, then increase $\epsilon
^*_j$ by a small quantity, say, $\zeta$.
As before, in our example, we chose $\zeta=10^{-10}$.
\item[(ii)] Calculate the probability of $\bigcup_{j=1}^kS_j$ as $P=\#\{
\bigcup_{j=1}^kS_j\}/N$.\vspace*{1pt}
\item[(iii)] Repeat steps (i) and (ii) until $P\approx0.95$ or any
desired probability.
\end{longlist}
Step (i) implicitly assumes that, since $C^{(i)}\notin S_j$, $S_j$ must
be a region with low probability,
so its expansion is necessary to increase the probability. This
expansion is achieved by increasing
$\epsilon^*_j$ by $\zeta$.
This step also ensures that the sets $S_j$ are selected adaptively, by
adaptively increasing $\epsilon^*_j$.
The final union of the $S_j$'s is the desired approximate HPD region.

\section{Nonuniqueness of clusterings and a suitable metric for comparison}
\label{sec:metric}

When we have two clusterings it is not very easy to compare them, as
the cluster labels of one
clustering may be quite unrelated to the cluster labels of the other.
One way to compare them
is to find a measure of divergence between them after permuting the
arbitrary indices to make
the two clusterings as close to each other as possible.

\ctn{Ghosh08} propose a simple way of capturing the similarity or
dissimilarity of two Clusterings $I$ and $\mathit{II}$ by setting up a
two-way table, where the frequency $n_{ij}$ in the $(i,j)$th cell is
the number of units belonging
to the $i$th cluster in $I$ and the $j$th cluster in $\mathit{II}$ [denoted
henceforth by $C_i(I)$ and $C_i(\mathit{II})$, resp.]. If two clusterings are very similar, the two-way table
will appear like a permutation
of rows and columns of a diagonal matrix with small perturbations.
For simplicity, we consider the case where the number of clusters is
the same for the two clusterings,
although the method can be extended easily to the case with an unequal
number of clusters. Suppose
that we fix the cluster numbers of $I$ and rename the clusters of~$\mathit{II}$
so as to make it most similar
to $I$. This means we try to rearrange the columns of the two-way table
so as to maximize the diagonal
elements of the table. We suggest that the larger the diagonal elements
(and hence the smaller the
off-diagonal ones), the closer the clusterings. Thus, a measure of
divergence may be based on the number
of units corresponding to the off-diagonal cells of the table.

\ctn{Ghosh08} define the distance $d(I,\mathit{II})$ between $I$ and $\mathit{II}$ as follows:
%
\begin{equation}
d(I,\mathit{II})=\min[n_{00}-(n_{1j_1}+n_{2j_2}+\cdots+n_{kj_k})]/n_{00}
\label{eq:metric}
\end{equation}
over all permutations $(j_1,j_2,\ldots,j_k)$ of $(1,2,\ldots,k)$,
where $k$ denotes the number of
clusters and $n_{00}=\sum\sum n_{ij}$ is the total number of units.

An upper bound for the metric $d(I,\mathit{II})$ for two Clusterings $I$ and
$\mathit{II}$, each with cardinality
$k$, and total number of units $n_{00}$, is given by $1-\frac
{mk}{n_{00}}$, where $m= [\frac{n_{00}}{k^2} ]$.
This upper bound is attained when $n_{ij}$'s in the two-way table are equal.

An alternative way to define the same distance is as follows. For each
unit, say, the $i$th unit, define
$S_i(I,\mathit{II})=0$ if the $i$th unit falls into $C_j(I)$ and $C_j(\mathit{II})$ for
the same $j$; otherwise set
$S_i(I,\mathit{II})=1$. Then $d(I,\mathit{II})$ defined earlier is the minimum value of
$\frac{\sum_{i=1}^{n_{00}}S_i(I,\mathit{II})}{n_{00}}$
over all possible numbering of the clusters of Clustering $\mathit{II}$.

If the number of clusters is not the same for the two partitions, one
may proceed as above with $\mathit{II}$
representing the partition with bigger cardinality. We would get the
same measure if we take the infimum
over all permutations of rows and columns. \ctn{Ghosh08} show that $d(I,\mathit{II})$
satisfies the properties of a metric.

\subsection{A simple approximation of the metric calculation}

It is, however, important to appreciate the fact that calculation of
the metric requires
taking the minima over all possible permutations of the clusters, and
for an even moderate number of clusters
this strategy leads to enormous computational burden. For MCMC samples,
one needs to compute
the metric for a~very large number of iterations, and since each
iteration may yield
at least a~moderate number of clusters, the calculation very quickly
becomes infeasible.
To rid the method of the
computational difficulty, we propose a simple heuristic approximation.

For any two reasonably close clusterings, after rearrangement, the
diagonal is likely to contain
the largest element. This suggests that, for such clusterings, in any
given column (or in any given row),
there is likely to be a single large element. Or, in other words, the
proportion of more than one
large element occurring in a single column (row) is negligible.
Formally, in such situations,
%
\begin{eqnarray}\label{eq:approx_metric}
d(I,\mathit{II})&=&\min[n_{00}-(n_{1j_1}+n_{2j_2}+\cdots
+n_{kj_k})]/n_{00}\nonumber\\
&\approx&\sum_{i=1}^k \Bigl\{n_{i0}-\max_{1\leq j\leq k}n_{ij}
\Bigr\} /n_{00}\\
&=&\tilde d(I,\mathit{II}).\nonumber
\end{eqnarray}
In the above, $n_{i0}=\sum_{j=1}^kn_{ij}$. Equation (\ref
{eq:approx_metric}) can be rewritten as
%
\begin{eqnarray}
\tilde d(I,\mathit{II})&=& \Biggl\{n_{00}-\sum_{i=1}^k\max_{1\leq j\leq
k}n_{ij} \Biggr\} \big/n_{00}\label{eq:approx1}\\
&=&1-\frac{\sum_{i=1}^k\max_{1\leq j\leq k}n_{ij}}{n_{00}}.\label
{eq:approx2}
\end{eqnarray}
Thus, (\ref{eq:approx2}) holds when the number of equalities among the
permutations $(j_1,j_2,\ldots,j_k)$
is negligible. Note, however, that (\ref{eq:approx2}) is not
symmetric, that is,
$\tilde d(I,\mathit{II})\neq\tilde d(\mathit{II},I)$; as a result, we symmetrize the
approximation by using
%
\begin{equation}
\hat d(I,\mathit{II})=\max \{\tilde d(I,\mathit{II}),\tilde d(\mathit{II},I) \}.
\label{eq:approx3}
\end{equation}
The reason for taking maximum, rather than other symmetrizing
transformations, such as average,
is that, if one of the two quantities $\tilde d(I,\mathit{II})$
or~$\tilde d(\mathit{II},I)$ is high, it indicates that the actual distance between the two
clusterings
cannot be small. Obviously, the aforementioned approximation is valid
even when the two Clusterings $I$ and $\mathit{II}$ consist of
a different number of clusters. It is also worth noting that $\hat d$
satisfies the first three properties of a metric, that is, $\hat
d(I,\mathit{II})\geq0$,
$\hat d$ is symmetric, and $\hat d(I,\mathit{II})=0$ if and only if $I$ and $\mathit{II}$
are equivalent in the sense that any one of $I$ and $\mathit{II}$ can be obtained
from the other by just a renumbering of the clusters. We prove these in
Section S-3. Although we have not been able to prove the fourth
property, that is, the triangular inequality
is satisfied by $\hat d$ in general, we have not been able to find a
counterexample to this effect, and, in fact, in all the examples we
have come across
the triangular inequality has been satisfied. Moreover, we prove in
Section S-4 that the triangular inequality holds when the clusterings
are independent in the sense that $n_{ij}=n_{i0}n_{0j}/n_{00}$, where\vspace*{-2pt}
$n_{i0}=\sum_jn_{ij}$, $n_{0j}=\sum_i n_{ij}$.
Hence, we \textit{conjecture} that $\hat d$ is also a metric. Also, the\vspace*{-2pt}
same upper bound
$1-\frac{mk}{n_{00}}$ as in the case of the metric $d$ is attained by
$\hat d$ as well when $n_{ij}$'s in the two-way layout are equal.
We demonstrate below with examples that the approximate metric (\ref
{eq:approx3})
agrees closely with the exact metric (\ref{eq:metric}).

\subsection{Illustration of the performance of the clustering metric
with simulated and real data}
\label{subsec:metric_performance}

In each of the examples illustrated below, we cluster the data into the
desired number of partitions
using the $K$-means algorithm, using two different starting points or
data sets with different sets of features.
This yields two clusterings
in each case, which we generically denote as Clustering~$I$ and
Clustering~$\mathit{II}$.

\subsubsection{Example 1: Performance of the cluster metric in the
case of simulated nonoverlapping clusters}
\label{subsubsec:example1}

We generate 5,000 observations from a mixture of five normal distributions
$N(i,\sigma^2)$, $i=1,\ldots,5$, with equal weights for specified
values of $\sigma$. This set of data
is then partitioned into 5 clusters with two different starting points
under the $K$-means algorithm,
yielding Clusterings $I$ and $\mathit{II}$.
The two clusterings, corresponding to the data generated with $\sigma
=0.25$, completely agree with each other,
and both $d$ and $\hat d=\max \{0,0 \}$ correctly yield the
value 0.

\begin{table}[b]
\caption{Two-way table showing number of observations in $C_i(I)\cap C_j(\mathit{II}),$ $i,j=1,\ldots,5$
for 5,000 observations drawn from the normal mixture $\sum_{i=1}^5\frac{1}{5}N(i,1)$}
\label{table:table2}
\begin{tabular*}{\tablewidth}{@{\extracolsep{\fill}}lcccccc@{}}
\hline
 & \multicolumn{5}{c}{\textbf{Clusters of Clustering} $\bolds{\mathit{II}}$}
\\[-5pt]
&\multicolumn{5}{c}{\hrulefill}\\
\multirow{2}{50pt}[11pt]{\textbf{Clusters of Clustering} $\bolds{I}$} & \textbf{1} & \textbf{2} & \textbf{3} & \textbf{4} & \textbf{5} & \textbf{Row sum}\\
\hline
1 & \hphantom{00}0 & \hphantom{000,}0 & \hphantom{000,}0 & \hphantom{00,}60 & 639 & \hphantom{0,}699\\
2 & \hphantom{00}0 & \hphantom{0,}229 & 1,086 & \hphantom{000,}0 & \hphantom{00}0 & 1,315\\
3 & 639 & \hphantom{000,}0 & \hphantom{000,}0 & \hphantom{000,}0 & \hphantom{00}0 & \hphantom{0,}639\\
4 & \hphantom{00}0 & \hphantom{000,}0 & \hphantom{0,}143 & 1,103 & \hphantom{00}0 & 1,246\\
5 & 166 & \hphantom{0,}935 & \hphantom{000,}0 & \hphantom{000,}0 & \hphantom{00}0 & 1,101\\[3pt]
Col. sum & 805 & 1,164 & 1,229 & 1,163 & 639 & 5,000\\
\hline
\end{tabular*}
\legend{Clusterings $I$ and $\mathit{II}$ are obtained~by~$K$-means clustering with two different starting points.}
\end{table}

\subsubsection{Example 2: Performance of the clustering metric in the
case of simulated overlapping clusters}
\label{subsubsec:example2}

We now give an example where the two clusterings are not exactly equal.
In this case we repeat Example 1, but with $\sigma=1$ instead of
$\sigma=0.25$.
Table \ref{table:table2} compares the two resulting clusterings.
In this case the clusterings are not equivalent, although there is a~one-to-one correspondence
between the two sets of clusters. For example, $C_1(\mathit{II})$ corresponds to
$C_3(I)$, but the 805 units of
$C_1(\mathit{II})$ are split into two parts---639 of them constitute the whole
of $C_3(I)$ and the
remaining 166 falls in~$C_5(I)$.
Here the distance $d$ between the two clusterings is given by $0.12$, while
the approximate metric $\hat d=\max \{0.12,0.1196 \}$
yields also exactly the same distance $0.12$.
Thus, in spite of the fact that the clusterings are not perfectly
equivalent, the approximate
metric $\hat d$ yields the exact answer.

\begin{table}
\tabcolsep=0pt
\caption{Two-way table showing number of units in $C_i(I)\cap C_j(\mathit{II}),$ $i,j=1,\ldots,11$
for the Western Ghats~data}
\label{table:table3}
\begin{tabular*}{\tablewidth}{@{\extracolsep{\fill}}ld{3.0}d{4.0}d{4.0}d{4.0}d{4.0}d{5.0}d{4.0}d{4.0}d{4.0}d{4.0}d{2.0}d{5.0}@{\hspace*{-3pt}}}
\hline
& \multicolumn{11}{c}{\textbf{Clusters of Clustering} $\bolds{\mathit{II}}$}\\[-5pt]
& \multicolumn{11}{c}{\hrulefill}\\
\multirow{2}{50pt}[11pt]{\textbf{Clusters of Clustering} $\bolds{I}$} & \multicolumn{1}{c}{\textbf{1}} & \multicolumn{1}{c}{\textbf{2}} &
\multicolumn{1}{c}{\textbf{3}} & \multicolumn{1}{c}{\textbf{4}} & \multicolumn{1}{c}{\textbf{5}} &
\multicolumn{1}{c}{\textbf{6}} & \multicolumn{1}{c}{\textbf{7}} & \multicolumn{1}{c}{\textbf{8}} &
\multicolumn{1}{c}{\textbf{9}} & \multicolumn{1}{c}{\textbf{10}} & \multicolumn{1}{c}{\textbf{11}} &
\multicolumn{1}{c@{\hspace*{-3pt}}}{\multirow{2}{18pt}[11pt]{\centering\textbf{Row sum}}}\\
\hline
\hphantom{0}1 & 0 & 0 & 0 & 0 & 0 & 0 & 0 & 0 & 0 & 0 & 2 & 2\\
\hphantom{0}2 & 0 & 0 & 0 & 0 & 0 & 0 & 0 & 0 & 886 & 57 & 0 & 943\\
\hphantom{0}3 & 0 & 2 & 0 & 0 & 0 & 0 & 711 & 1\mbox{,}432 & 1\mbox{,}940 & 15 & 0 & 4\mbox{,}100\\
\hphantom{0}4 & 0 & 0 & 0 & 0 & 0 & 0 & 0 & 0 & 0 & 0 & 48 & 48\\
\hphantom{0}5 & 0 & 3 & 0 & 1\mbox{,}781 & 0 & 0 & 86 & 0 & 2 & 0 & 0 & 1\mbox{,}872\\
\hphantom{0}6 & 0 & 0 & 0 & 0 & 0 & 0 & 0 & 0 & 2 & 0 & 0 & 2\\
\hphantom{0}7 & 0 & 198 & 1\mbox{,}076 & 86 & 77 & 0 & 6\mbox{,}053 & 1\mbox{,}877 & 0 & 0 & 0 & 9\mbox{,}367\\
\hphantom{0}8 & 0 & 516 & 6\mbox{,}859 & 4\mbox{,}630 & 3\mbox{,}683 & 0 & 2 & 0 & 0 & 0 & 0 & 15\mbox{,}690\\
\hphantom{0}9 & 182 & 5 & 0 & 0 & 0 & 102 & 0 & 474 & 0 & 1\mbox{,}920 & 0 & 2\mbox{,}683\\
10 & 502 & 317 & 0 & 0 & 5\mbox{,}686 & 10\mbox{,}271 & 0 & 127 & 0 & 0 & 0 & 16\mbox{,}903\\
11 & 214 & 2 & 0 & 1 & 0 & 0 & 0 & 0 & 0 & 0 & 7 & 224\\[3pt]
Col. sum & 898 & 1\mbox{,}043 & 7\mbox{,}935 & 6\mbox{,}498 & 9\mbox{,}446 & 10\mbox{,}373 & 6\mbox{,}852 & 3\mbox{,}910 & 2\mbox{,}830 & 1\mbox{,}992 & 57 & 51\mbox{,}834\\
\hline
\end{tabular*}
\legend{Row-wise clusters correspond to Clustering $I$
and column-wise clusters correspond to Clustering $\mathit{II}$.
Clusterings $I$ and $\mathit{II}$ are obtained by $K$-means clustering with two different starting points.}
\vspace*{-2pt}
\end{table}

\vspace*{-2pt}
\subsubsection{Example 3: Performance of the clustering metric in the
case of real data}\label{subsubsec:example3}

We now consider the real data obtained from the Western
Ghats. The data consist of multivariate (4-variate) observations
related to vegetation indices for
51,834 ``super pixels'' throughout the Western Ghats region obtained
from the imagery generated
by Indian remote sensing satellites. We do the clustering with a number
of clusters fixed at 11 as finally obtained
in~NG. Table \ref{table:table3} provides a comparison between
Clusterings $I$ and $\mathit{II}$
(obtained from two different sets of initial values of the $K$-means
clustering algorithm). There is no obvious one-to-one
correspondence between the clusters of the two clusterings. For
example, cluster $C_8(I)$ is split
into three large parts of sizes 6,859, 4,630 and 3,683 which correspond to
$C_3(\mathit{II}), C_4(\mathit{II})$
and~$C_5(\mathit{II})$, respectively. The distance $d$ between the two clusterings
in this case turns out to be $0.432$, whereas $\hat d=\max \{
0.42169,0.22248 \}$ yields 0.422. This difference is obviously
due to the lack of one-to-one correspondence between the clusters;
however, the approximation
is still quite accurate.

\vspace*{-2pt}
\subsubsection{Example 4: Performance of clustering metric and the
effect of addition or
deletion of a variable in the multivariate case}
\label{subsubsec:example4}

The Western Ghats data consist of 4-variate observations for 51,834
cases (units).
We wish to study the change in the clusterings if a variable is added
or deleted. Table~\ref{table:table4}
provides a comparison between Clustering $I$ obtained using the
$K$-means algorithm and three of the variables,
while Clustering $\mathit{II}$ is obtained using the $K$-means algorithm
and all the four variables.\vadjust{\eject} It is expected that a cluster in Clustering
$I$ will be split into
more than one cluster of Clustering $\mathit{II}$ where the additional
information on the 4th variable is used.
On the other hand, some of the clusters in Clustering $\mathit{II}$ are expected
to coalesce when the 4th variable
is dropped. In Table \ref{table:table4}, however, we observe split in
both the directions. This is because
we are fixing the same number of clusters in both Clustering $I$ (with
three variables) and Clustering $\mathit{II}$
(with four variables). In this case, however, the value of the exact
distance metric $d$ is 0.283, while
the approximated value obtained using $\hat d=\max \{
0.10837,0.28211 \} $ is 0.282, again exhibiting
quite accurate approximation.

\begin{table}
\tabcolsep=0pt
\caption{Two-way table showing number of units in $C_i(I)\cap C_j(\mathit{II}),$ $i,j=1,\ldots,11$
for the Western Ghats~data}
\label{table:table4}
\begin{tabular*}{\tablewidth}{@{\extracolsep{\fill}}lcd{3.0}d{4.0}d{2.0}d{4.0}cd{4.0}d{5.0}d{4.0}d{5.0}d{3.0}d{5.0}@{\hspace*{-3pt}}}
\hline
& \multicolumn{11}{c}{\textbf{Clusters of Clustering $\bolds{\mathit{II}}$}}\\[-5pt]
& \multicolumn{11}{c}{\hrulefill}\\
\multirow{2}{50pt}[11pt]{\textbf{Clusters of Clustering} $\bolds{\mathit{I}}$} & \textbf{1} & \multicolumn{1}{c}{\textbf{2}} &
\multicolumn{1}{c}{\textbf{3}} & \multicolumn{1}{c}{\textbf{4}} & \multicolumn{1}{c}{\textbf{5}} &
\textbf{6} & \multicolumn{1}{c}{\textbf{7}} & \multicolumn{1}{c}{\textbf{8}} & \multicolumn{1}{c}{\textbf{9}} &
\multicolumn{1}{c}{\textbf{10}} & \multicolumn{1}{c}{\textbf{11}} &
\multicolumn{1}{c@{\hspace*{-3pt}}}{\multirow{2}{18pt}[11pt]{\centering\textbf{Row sum}}}\\
\hline
\hphantom{0}1 & 0 & 0 & 0 & 0 & 0 & 2 & 0 & 0 & 0 & 0 & 0 & 2\\
\hphantom{0}2 & 0 & 929 & 158 & 0 & 2 & 0 & 0 & 0 & 1 & 0 & 0 & 1\mbox{,}090\\
\hphantom{0}3 & 0 & 0 & 3\mbox{,}814 & 0 & 6 & 0 & 252 & 0 & 0 & 0 & 0 & 4\mbox{,}072\\
\hphantom{0}4 & 0 & 0 & 0 & 39 & 1\mbox{,}796 & 0 & 78 & 1\mbox{,}085 & 0 & 0 & 1 & 2\mbox{,}999\\
\hphantom{0}5 & 0 & 0 & 0 & 0 & 23 & 0 & 8\mbox{,}663 & 3 & 0 & 0 & 1 & 8\mbox{,}690\\
\hphantom{0}6 & 0 & 0 & 0 & 0 & 0 & 0 & 0 & 0 & 197 & 4\mbox{,}067 & 45 & 4\mbox{,}309\\
\hphantom{0}7 & 0 & 0 & 0 & 0 & 41 & 0 & 44 & 9\mbox{,}622 & 0 & 0 & 1 & 9\mbox{,}708\\
\hphantom{0}8 & 0 & 14 & 128 & 0 & 0 & 0 & 49 & 0 & 2\mbox{,}451 & 30 & 7 & 2\mbox{,}679\\
\hphantom{0}9 & 0 & 0 & 0 & 0 & 0 & 0 & 0 & 0 & 1 & 9\mbox{,}737 & 0 & 9\mbox{,}738\\
10 & 2 & 0 & 0 & 9 & 0 & 0 & 0 & 0 & 33 & 13 & 156 & 213\\
11 & 0 & 0 & 0 & 0 & 4 & 0 & 281 & 4\mbox{,}980 & 0 & 3\mbox{,}056 & 13 & 8\mbox{,}334\\[3pt]
Col. sum & 2 & 943  & 4\mbox{,}100 & 48 & 1\mbox{,}872 & 2 & 9\mbox{,}367 & 15\mbox{,}690 & 2\mbox{,}683 & 16\mbox{,}903 & 224 & 51\mbox{,}834\\
\hline
\end{tabular*}
\legend{Row-wise clusters correspond to Clustering $I$
with three variables and column-wise clusters correspond to Clustering $\mathit{II}$ with four variables.
Clusterings $I$ and $\mathit{II}$ are obtained by $K$-means clustering.}
\vspace*{-2pt}
\end{table}

\vspace*{-2pt}\section{Application to the Western Ghats data}\vspace*{-2pt}
\label{sec:real_data}

\subsection{Data description}
\label{subsec:data_description}

NG [see also \ctn{Nagendra99}] consider a broad scale mapping of the
Western Ghats into different
landscape types based on satellite imagery, using the Normalized
Difference Vegetation Index (NDVI). The index
is believed to be correlated to vegetation biomass, vigour,
photosynthetic activity
and leaf area index, and is known to be potentially useful for
classifying different
types of vegetation.
Another important advantage of NDVI is that it reduces problems of
scene-to-scene radiometric variability
of the remotely sensed satellite images.
For each $50\times50$ pixel unit (the resolution being
$36.5\times36.5$~m for each of the 2,500 pixels) constituting a
``superpixel,''
the four moments of distribution---mean, standard deviation, skewness
and kurtosis,
were calculated.\vadjust{\eject} These superpixels were then clustered using
unsupervised classification;
NG report the final number of clusters to be 11.
The distribution of
the clusters are to be interpreted in terms of topography, climate,
population, agriculture
and vegetation cover. For further details regarding the data and the
methodology,
we refer the reader to NG.

\begin{figure}

\includegraphics{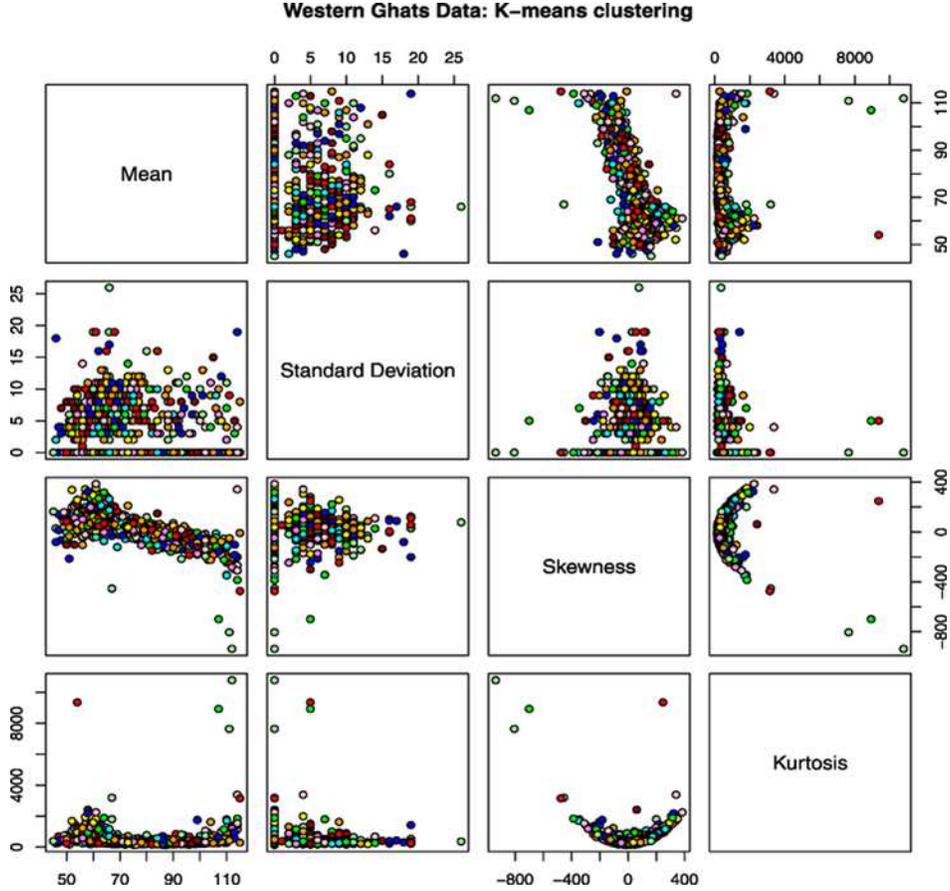}

\caption{Pairwise scatterplots of the 4-variables used for clustering the Western Ghats data.
Different colours denote different clusters corresponding to the $K$-means clustering shown
in Figure~\protect\ref{fig:kmeans}.}
\label{fig:scatterplots}
\end{figure}

\begin{figure}

\includegraphics{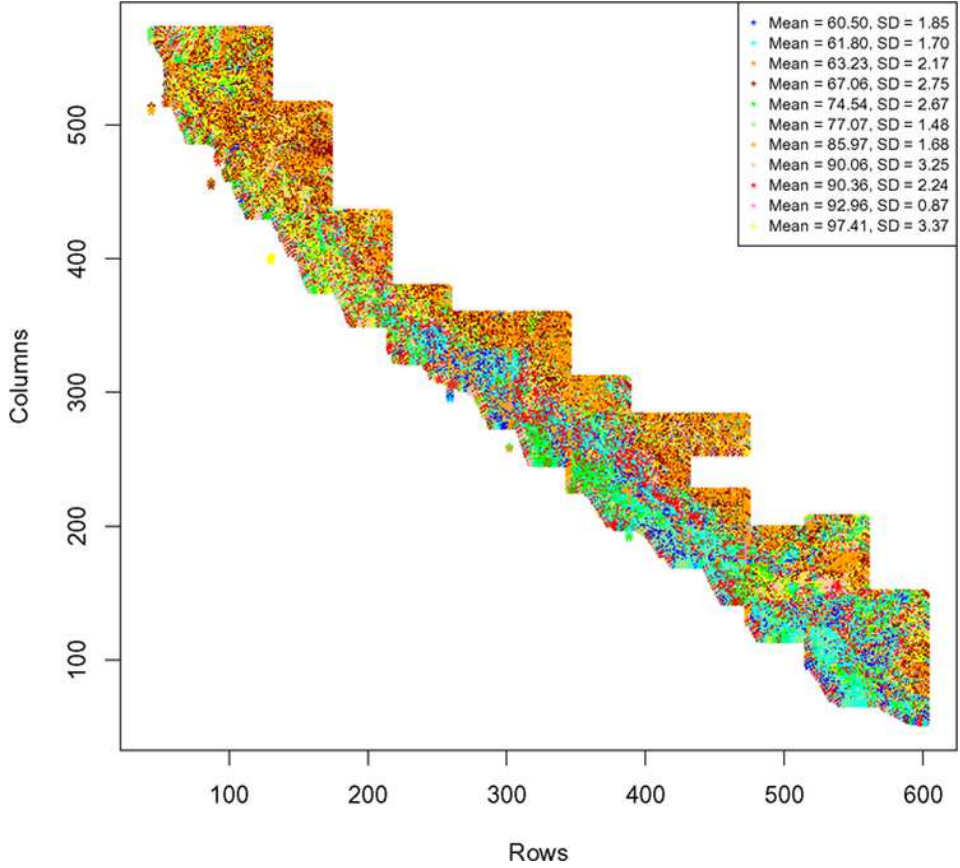}

\vspace*{-2pt}
\caption{$K$-means clustering; different colours denote 11 different
clusters.
Cluster averages of mean (Mean) and standard deviation (SD) are shown in the legend.}
\vspace*{-2pt}
\label{fig:kmeans}
\end{figure}

The pairwise scatterplots of the four variables used for clustering the
Western Ghats data
is shown in Figure \ref{fig:scatterplots}.
Only for this plotting purpose, the data set is thinned to include 1
four-variate observation
in every 50 such observations.
The data points within the scatterplots are colored
differently to show 11 different clusters, obtained using $K$-means
clustering, a proxy for the method
used by NG for their clustering. The $K$-means clustering, which has
been analyzed in detail in subsequent
subsections, is displayed in Figure~\ref{fig:kmeans}. Each point in
the latter figure corresponds to a 4-variate
observation indexed by its
position of the form $(i,j)$, where $i$ and $j$ represent the spatial
coordinates, namely,
row and column numbers, respectively, on a relevant
spatial grid.

It is important to note that NG has ignored the spatial structure of
the superpixels while analyzing
the data. It was perhaps anticipated by NG that the clustering would
not change nonnegligibly
by incorporating the spatial locations because of the huge and quite
informative data. The
computational difficulties associated with spatial methods with data
sets as huge as this may be another
quite pragmatic reason.
But whatever the reasons of NG, it is perhaps worth investigating
statistically, whether or not omission
of the spatial structure is inconsequential.
To this end, we carried out a simple, informal test, reported in
Section S-5. Since the test indicated
insignificance of the spatial structure, we proceeded with the same
data set used by NG.

\subsection{Choice of prior}
\label{subsec:prior}
We chose $\bmu_0$ and $\bS$ to be the mean and the covariance matrix
of the data, respectively, $s=4$, the minimum degrees
of freedom required to make $\bG_0$ well-defined, and $\psi=1$. These
choices are natural, and in this Western Ghats example, with massive data,
robustness of the priors is ensured.
But appropriate choices of $M$ and the prior of $\alpha$ are
important, and here we have been
guided by the results obtained by NG.
For instance, the final clustering obtained by NG, with their method
that uses the $K$-means method and a subjective merging
procedure, consists of 11 clusters. However, they initially started
with 20 clusters, obtaining 11 clusters finally.
In our model, we set $M=30$ to account for some extra uncertainty. In
fact, a maximum of 30 components has also been used by \ctn
{Richardson97} and SB.
For the scale parameter $\alpha$ we considered the prior $\alpha\sim
\operatorname{Gamma}(0.1,0.1)$, that is, a Gamma distribution with mean 1 and
variance 10.
This prior is reasonably close to noninformative, and, importantly,
with these prior choices, 11 clusters get the maximum posterior mass,
matching the
number of clusters obtained by NG. A detailed study of sensitivity of
the posterior inference with respect to other choices of the priors is
reported in Section~S-6.

\subsection{Gibbs sampling for computing the posterior distribution of
clustering}
\label{subsec:gibbs_sampling}

Apparently, for our purpose, the marginalized version of SB's model
described in Section \ref{sec:dp}
seems preferable since here we are only interested in the posterior
distribution of clustering,
and hence retaining the parameters~$\bTheta_M$ seems to serve no
purpose. However, the expressions in
Sections~\mbox{S-1} and S-2
show that calculation of the full conditional probabilities of $z_i$
in the marginalized version involves much more computational complexity
compared to the nonmarginalized version. Since these
computational complexities are multiplied $n$ times while updating the
complete $\bZ$ vector, with $n=51\mbox{,}834$, the marginalized version
tends to be infeasible for massive data. Indeed, for the marginalized
version, it took about 30 hours to complete just 10 iterations.
We remark that implementation of EW's model using the marginalized
algorithm proposed in \ctn{MacEachern94} took more
than 39 hours to generate just 10 MCMC realizations.
On the other hand, for the nonmarginalized version of SB's model,
generation of 30,000 MCMC samples,
which includes a burn-in of 10,000, took just around 14 hours. In
Section S-7 we provide a thorough account of the computational superiority
of SB's model compared to that of EW. Section S-8 provides a new method
based on clusterings to assess convergence of our
Gibbs sampler. Excellent convergence is indicated by this methodology.

\subsection{Posterior distribution of the number of clusters}
\label{subsec:posterior_no_clusters}

The posterior probabilities of the number of components being $\{6,
\ldots, 18\}$
are $\{0.00025, 0.00395,\break 0.02955, \ 0.10600,
\ 0.20815, \ 0.25135, \ 0.20715, \ 0.12190, \ 0.05205, \ 0.01555,\break 0.00345, 0.00055,
0.00010\}$, respectively,
while the other values have zero posterior probabilities. Thus, 11
components have the maximum posterior
probability, 0.25135. The components in this example all turned out to
be nonempty, which is to be expected
since the data set is so large. Even with other experiments with this
data set, using SB's model, this same
fact was observed.
Hence, we will use the terms ``clusters'' and ``components''
interchangeably from this point on.
It is striking to note that NG also obtained 11 clusters with their
analysis of the
Western Ghats data.

\subsection{Bayesian central clustering of the Western Ghats data}
\label{subsec:data_central_clustering}

\begin{figure}

\includegraphics{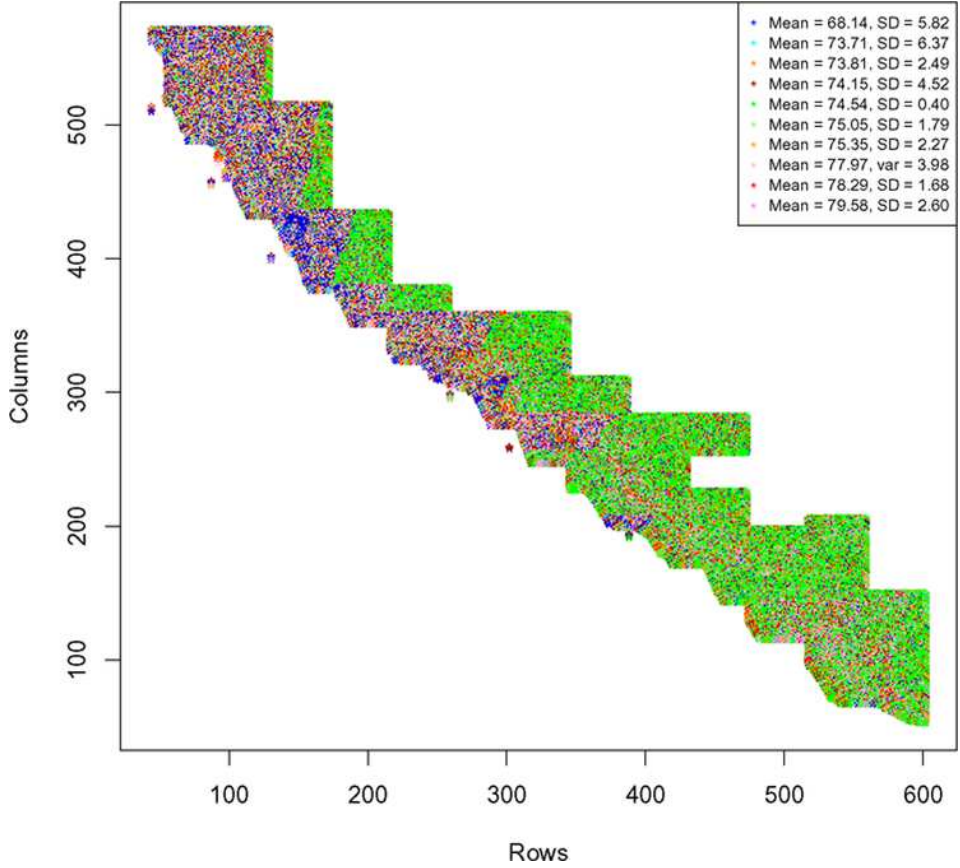}%
\caption{Modal central clustering $C^{(412)}$; different colours denote 10 different
clusters.
Cluster averages of mean (Mean) and standard deviation (SD) are shown in the legend.}
\label{fig:clustering412}
\end{figure}

\begin{figure}

\includegraphics{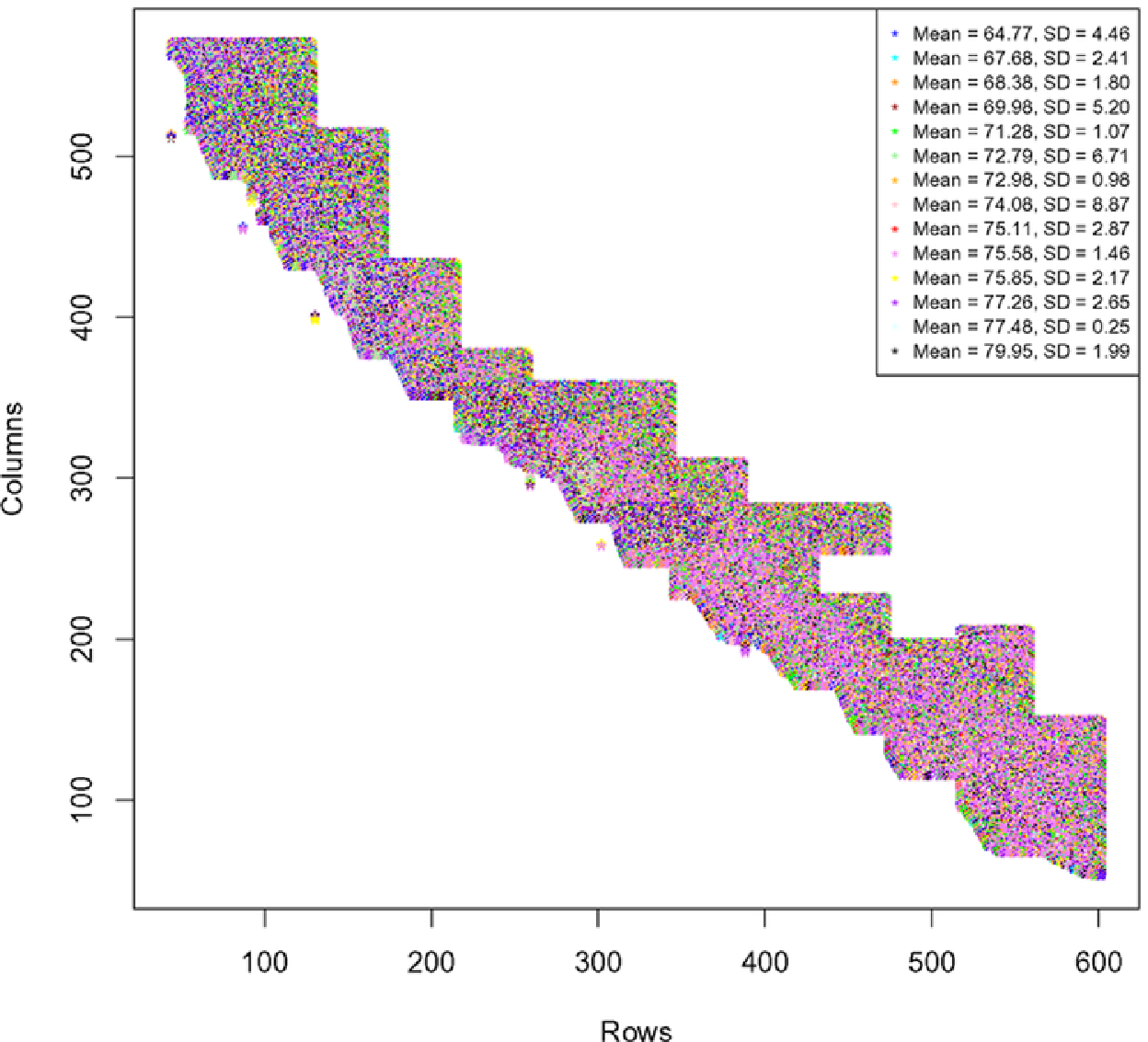}%
\caption{Modal central clustering $C^{(759)}$; different colours denote 14 different
clusters.
Cluster averages of mean (Mean) and standard deviation (SD) are shown in the legend.}
\label{fig:clustering759}
\end{figure}

We obtained two central clusterings: the one obtained in the 759th
iteration, corresponding
to $\epsilon=0.65$, which consists of 14 clusters, and another
obtained in the 412th iteration, corresponding to $\epsilon=0.70$,
consisting of 10 clusters.
It is worth noting that the empirical probabilities
$\frac{1}{N}\# \{C^{(k)};1\leq k\leq N\dvtx\hat
d(C^{(i)},\break C^{(k)})<\epsilon \}$
for $i=1,\ldots,N$, turned out to be zero
for $\epsilon<0.65$. For $\epsilon>0.70$ both clusterings
corresponding to the 412th and the 759th iterations
maximized the aforementioned empirical probabilities. Hence, the
clustering corresponding to the 759th iteration is an
estimate of the global mode
of the posterior of clustering as it corresponds to the smaller
$\epsilon$.
Figure \ref{fig:clustering412} shows the clustering of the modal clustering,
$C^{(412)}$. The other clustering, $C^{(759)}$, is displayed in Figure
\ref{fig:clustering759}.

The two modal clusterings are close to each other, the distance being
0.649, even though the number
of their clusters differ. Although one might suspect, because of the
relative closeness of the two modes,
that some clusters of the 10-cluster mode $C^{(412)}$ are simply split
up to give rise to the 14-cluster
mode $C^{(759)}$, this is not the case, as is also evident from the
average means and average standard deviations
reported in the legends of Figures \ref{fig:clustering412} and~\ref
{fig:clustering759}.
The average means and the average standard deviations of at least some
clusterings would have been the same across
the two figures had this been the case.

It is not surprising that the two
central clusterings consist of
14 and 10 clusters, although 11 clusters have the maximum posterior
probability. This is because the
Bayesian central clustering has been obtained unconditionally,
marginalizing over the number of components,
without fixing the number of components at 11. This issue, concerning
conditional and unconditional clusterings,
will be discussed in detail in Section \ref{subsec:conditional_clustering}.
Here we only note that the distance between two clusterings
need not be small even if the number of clusters are the same (see the examples
in Section \ref{subsec:metric_performance}); had this been the case,
conditional and unconditional clusterings would be the same.

\subsection{Bayesian 95\% credible and HPD regions}
\label{subsec:data_HPD}

Furthermore, with $\epsilon^*= 0.707$ and $\epsilon^*= 0.746$, we
obtained approximately 95\%
credible regions corresponding to the central clusterings $C^{(412)}$
and $C^{(759)}$, respectively. In both cases
the probability of the credible region turned out to be 0.951.\vadjust{\eject}
Since the distance between $C^{(412)}$ and $C^{(759)}$ is 0.649, each
falls within the 95\% credible
region of the other.
We also constructed the 95\% HPD region
using the two central clusterings. Employing the adaptive algorithm
provided in Section \ref{subsec:hpd},
we obtained $\epsilon^*_1=0.688$ and $\epsilon^*_2=0.710$
corresponding to $C^{(412)}$ and $C^{(759)}$, respectively.
The probability of the HPD region is 0.952.

\begin{figure}

\includegraphics{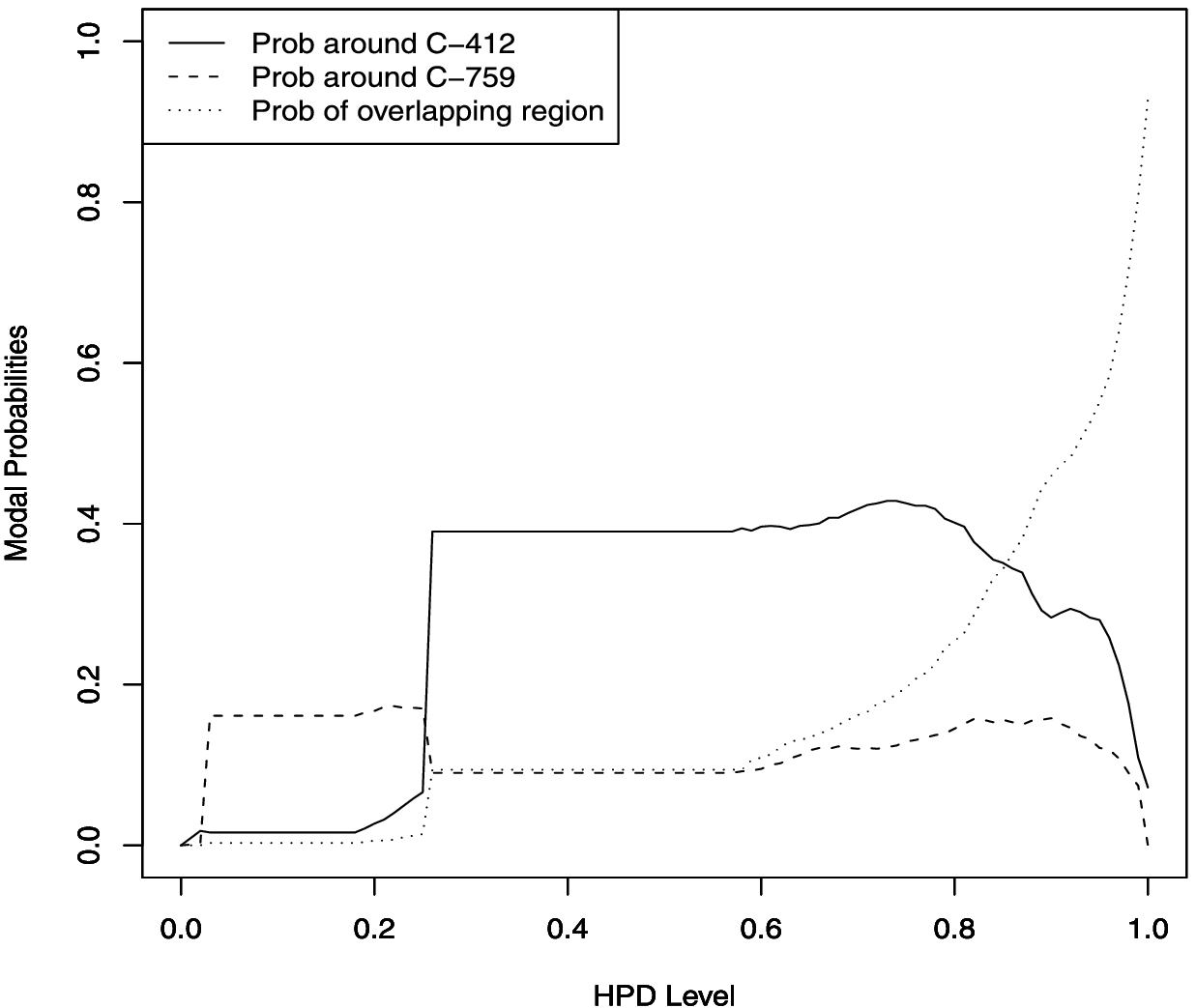}

\vspace*{-3pt}
\caption{Plots of probabilities around each of the two modes $C^{(412)}$ and $C^{(759)}$ against the corresponding
HPD levels. These probabilities exclude the probabilities of the intersection of the two modal regions
given by
$\operatorname{Pr}(\{C\dvtx\hat d(C^{(412)},C)<\epsilon_1\}\cap\{C\dvtx\hat d(C^{(759)},C)<\epsilon_2\})$, the values of which are plotted
separately against the corresponding HPD levels for appropriate values of $\epsilon_1$ and $\epsilon_2$.}
\label{fig:hpd_sequence}
\vspace*{-4pt}
\end{figure}

Figure \ref{fig:hpd_sequence} shows the probabilities around each of
the two modal clusterings (excluding
the probabilities of the overlapping regions) for different
levels of HPD. The probabilities of the overlapping regions for
different levels of HPD are also shown in the same
figure. Initially, that is, when the HPD levels were less than 0.3,
the probabilities around $C^{(759)}$ were greater than those around
$C^{(412)}$, but
from that point on the modal probabilities of $C^{(412)}$ were greater.
This is not surprising, since $C^{(759)}$ is the global mode
(see Section~\ref{subsec:data_central_clustering}) implying that for smaller
HPD levels most probability mass will be concentrated around its modal region.
But since its modal region must be smaller compared to that of
$C^{(412)}$, which is the local mode,
for higher HPD levels the former can accommodate only a small portion
of the entire HPD level. The remaining,
larger portion of the HPD level must be associated with the modal
region of the local mode.
Also, as to be expected, the probabilities
of the overlapping regions increased steadily with the HPD
levels.\looseness=-1

The distribution of the number of clusters of the clusterings
falling\break within the 95\% HPD regions are
as follows:
the number of clusters getting nonzero probabilities are $\{7, \ldots, 16\}$,
and their respective probabilities are
$\{0.004201681,\ 0.024159664,\ 0.101890756,\ 0.198529412,\ 0.255252101,\break
0.222689076,\ 0.122899160,\ 0.056722689,\ 0.009453782,\ 0.004201681\},$ showing that 11
clusters again receives
the maximum probability.

\subsection{Method of NG}
\label{subsec:NG_method}

NG essentially used a $K$-means clustering
algorithm [see, e.g., \ctn{Hartigan75}], fixing the number of
clusters to be 20.
Next, 14 clusters were obtained by merging some of the final 20 clusters.
These were further reduced to 11 clusters, the merging operation
justified on ecological grounds,
rather than classical statistical theory of clustering. We interpret
this ``ecological justification of merging''
as implicit use of subjective prior information.
Since the numerical results or the exact methodological steps of NG are
not available to us,
we used the $K$-means algorithm
with the number of clusters set equal to 11, as a proxy for the
methodology of NG.
Figure \ref{fig:kmeans} displays the $K$-means clustering of the
Western Ghats data set.
We, however, found that
the distance from the $K$-means
clustering to $C^{(759)}$ and~$C^{(412)}$ are 0.832 and 0.848,
respectively, signifying that
the $K$-means clustering does not fall within the 95\% credible or HPD
regions corresponding to our Bayesian
methodologies.
The reasons for this discrepancy between our Bayesian central
clustering and the $K$-means clustering
are discussed in detail in Section~\ref{subsec:conditional_clustering}.\vadjust{\eject}

\subsection{Bayesian conditional and unconditional central clusterings
and comparison with $K$-means
clustering}
\label{subsec:conditional_clustering}

The issue of conditioning of our Bayesian central clustering on $k$
clusters is the key
to understanding the discrepancy between central clustering and
$K$-means clustering,
which we now discuss.

\begin{figure}

\includegraphics{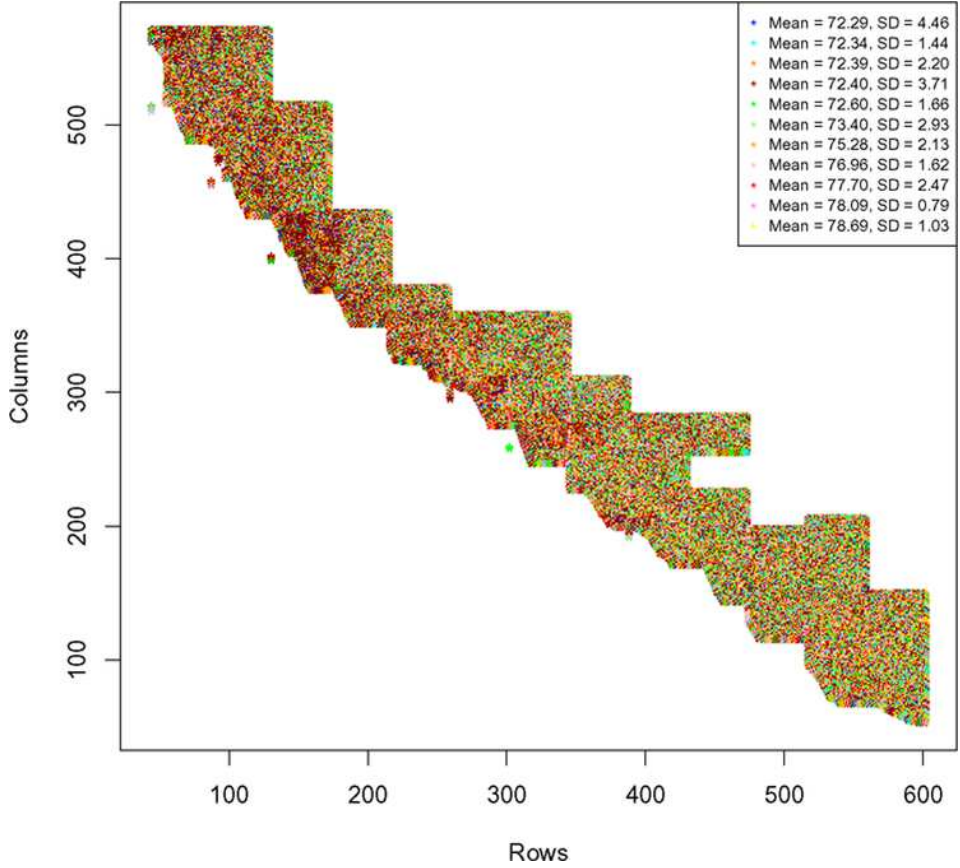}

\vspace*{-4pt}
\caption{Modal conditional central clustering $C^{(1)}$; different colours denote 11 different
clusters.
The first component of each of the distinct mean values $\{\mu^*_{1j};j=1,\ldots,11\}$, associated with the clusters,
are shown in the legend.}
\vspace*{-10pt}\label{fig:clustering_conditional}
\end{figure}

Our Bayesian method obtains central clustering by taking account of
uncertainties about the
number of clusters, while the $K$-means algorithm keeps the number of
clusters fixed, thus failing,
while clustering the data, to take account of the uncertainty involved
in clustering.
To vindicate this, we obtained Bayesian central clustering conditional
on 11 components.
The clustering in the very first iteration, denoted by $C^{(1)}$, now
turned out to be the central
clustering, and it remained so for all
$\epsilon\geq0.75$. For $\epsilon< 0.75$
for any $i\in\{1,\ldots,N\}$, the empirical probabilities
$\frac{1}{N}\# \{C^{(k)};1\leq k\leq N:\hat
d(C^{(i)},C^{(k)})<\epsilon \}$
turned out to be zero, suggesting that $C^{(1)}$ is the global mode,
conditional on 11 clusters.
The conditional central clustering $C^{(1)}$ is shown in
Figure~\ref{fig:clustering_conditional}.
The conditional 95\% credible region, which is also the conditional
95\% HPD region
because of unimodality,
is given by $\{C:\hat d(C^{(1)},C)<0.827\}$, for~tho\-se~$C$ having
11 clusters. The empirical probability
of this set is 0.95, indicating very good approximation to the true
credible region.
Importantly, the $K$-means clustering now falls within this 95\%
credible region, the
distance between $C^{(1)}$ and $K$-means clustering being 0.729.
We remark in this context that the distance between the central
clustering conditional on $k$ clusters
and the $K$-means clustering with 11 clusters is minimized when $k=11$.
That the unconditional 95\% Bayesian credible region does not include
the $K$-means clustering
but this conditional 95\% Bayesian credible region does shows that
$K$-means clustering fails
to account for the uncertainty in the number of clusters, even if one
fixes the number of clusters
very accurately in the $K$-means algorithm.

Hence, although the results of NG are not available to us,
we can conclude, based on our analyses, that the clustering they
obtained is unlikely to fall within our
unconditional 95\% credible or HPD regions,
even though broadly their clustering, plotted as Figure 1 in NG, looks
similar to our
Figure \ref{fig:clustering759}.
Their results are rather comparable with our conditional clustering,
shown in
Figure \ref{fig:clustering_conditional}. Detailed interpretation of
the results
and their comparisons are provided in Section \ref{sec:interpretation}.

\vspace*{-2pt}
\section{Detailed interpretation of the results of the Western Ghats
data analysis}
\label{sec:interpretation}
Following NG, we order the landscape types of Western Ghats in
ascending order of the means (the first component
of the 4-variate data vectors) within each cluster.
Since the clusterings obtained by us need not match that obtained by
NG, our ordering of the landscape types
need not agree with that of NG. But in spite of this,
the similarities between the clustering obtained by NG and our
$K$-means clustering seem to be substantial. Details
of landscape types and their comparisons with respect to different
clusterings are presented below.

\subsection{Landscape type-1}
\label{subsec:land1}

\vspace*{-4pt}
\subsubsection{Distribution in $K$-means clustering}
\label{subsubsec:kmeans_1}
Landscape type-1 of our $K$-means clustering (Figure \ref{fig:kmeans})
is distributed mainly in the south-east, and toward the middle part
of Western Ghats. Comparatively small parts of landscape 1 are also
distributed in the south-west region, and are
almost absent in the northern region. Fair amount of heterogeneity in
this landscape type is indicated by
the average standard deviation associated with this cluster. This shows
that this landscape comprises a mixture
of several ecosystems. From the description provided by NG about this
part of Western Ghats
(the location of landscape 1 of $K$-means clustering seems to
correspond to the locations of landscapes 1 and 2 of NG),
we can infer
that the natural vegetation of the south-east part of this landscape
area is tropical dry deciduous forest,\vadjust{\eject} where rice, millets and oilseeds
are grown. The middle part of the Ghats where this landscape is also
found comprises tropical moist deciduous forest.
Large parts of this landscape have been converted to open areas with
palmyra trees planted in between.
The small south-western parts of this landscape consist of moist
deciduous vegetation.

\vspace*{-1pt}
\subsubsection{Distribution in conditional clustering}
\label{subsubsec:conditional_1}
Landscape 1 of conditional clustering (Figure \ref
{fig:clustering_conditional}) is distributed all over
Western Ghats (corresponding to either of the similar landscape types
4, 5, 6 of NG), and the high standard deviation indicates the
very substantial number of ecosystems
it comprises. Natural vegetation is mostly dry deciduous in the north
and moist deciduous in the south.
The forests of the north have been replaced by tree savanna, shrub
savanna and open land complexes.
The south consists of open lands and palmyra trees.
Rice, millets and
oilseeds are planted in some parts of this landscape. The eastern parts
are of the montane wet evergreen forest type.\looseness=-1

\vspace*{-1pt}
\subsubsection{Distribution in clustering $C^{(412)}$}
\label{subsubsec:412_1}
In the case of $C^{(412)}$ (Figure \ref{fig:clustering412}), landscape
type 1 is distributed over the north-west
part of the Ghats, and is absent elsewhere.
High average standard deviation suggests that this landscape is a
mixture of many individual landscape elements.
This part is characterized by dry deciduous vegetation. As opposed to
the previous clusterings, in this case
landscape 1 does not seem to be consistent with any of the landscapes
of NG.

\vspace*{-1pt}
\subsubsection{Distribution in clustering $C^{(759)}$}
\label{subsubsec:759_1}
Consistent with the case of $C^{(412)}$, here also landscape 1 is
distributed mainly over the north-western part
of Western Ghats, and here also the average standard deviation is quite
high. Once again, this landscape is
not consistent with any landscape obtained by~NG.

\vspace*{-1pt}
\subsection{Landscape type-2}
\label{subsec:land_2}
\vspace*{-1pt}
\subsubsection{Distribution in $K$-means clustering}
\label{subsubsec:kmeans_2}
The distribution of landscape type~2 for $K$-means clustering is
similar to that of landscape type 1.
The average standard deviation is also comparable, and is only slightly less.

\vspace*{-1pt}
\subsubsection{Distribution in conditional clustering}
\label{subsubsec:conditional_2}
The distribution in this case is comparable to that of landscape 1 of
conditional clustering, only the variability
is much less, suggesting that fewer ecosystems have comprised this landscape.

\vspace*{-1pt}
\subsubsection{Distribution in clustering $C^{(412)}$}
\label{subsubsec:412_2}
In the case of $C^{(412)}$, landscape~2 is distributed mainly along the
north-western part, stretching
along the mid-western part of the Ghats, and also comprising some part
of the south-eastern part.
The large
variability suggests abundance of individual landscape elements. The
natural vegetation here is dry deciduous
forests in the north and moist deciduous forests toward the south.

\subsubsection{Distribution in clustering $C^{(759)}$}
\label{subsubsec:759_2}
Landscape 2 for $C^{(759)}$ stretches mainly from the middle part of
the Western Ghats extending till south,
where it is more prominent. Here also the variability is significant,
although smaller compared to that
of $C^{(412)}$. The vegetation here is mainly tropical and moist
deciduous forests.

\vspace*{-1pt}
\subsection{Landscape type-3}
\label{subsec:land_3}
\vspace*{-1pt}
\subsubsection{Distribution in $K$-means clustering}
\label{subsubsec:kmeans_3}
Landscape type 3, as also in the case of landscape type 3 of NG, is
present mainly along the eastern sides of Western Ghats.
The natural vegetation is of the montane wet evergreen and moist
deciduous forest type, and rice, millets and oilseeds
are grown.

\vspace*{-1pt}
\subsubsection{Distribution in conditional clustering}
\label{subsubsec:conditional_3}
With respect to the conditional clustering, landscape type 3 is
distributed along the entire length of the Western Ghats,
not mainly toward the eastern part as in the case of $K$-means clustering.

\vspace*{-1pt}
\subsubsection{Distribution in clustering $C^{(412)}$}
\label{subsubsec:412_3}
With respect to $C^{(412)}$, this landscape is distributed mainly
toward the eastern part of the Ghats, but also
generally along the entire region.

\vspace*{-1pt}
\subsubsection{Distribution in clustering $C^{(759)}$}
\label{subsubsec:759_3}
As in the previous clusterings, landscape 3 is distributed mainly along
the eastern side of the Ghats with respect
to $C^{(759)}$. The variability in this case is a little less than in
the case of other clusterings.

\vspace*{-1pt}
\subsection{Landscape type-4}
\label{subsec:land_4}
\vspace*{-1pt}
\subsubsection{Distribution in $K$-means clustering}
\label{subsubsec:kmeans_4}
As in the case of corresponding landscape 3, landscape 4 for $K$-means
is also distributed mainly toward the eastern region,
and in the northern part it is distributed in both eastern and western
parts, showing similarity with the
distribution of landscape 4 of NG. The variability is large enough to
suggest prevalence of a number of different ecosystems.

\vspace*{-1pt}
\subsubsection{Distribution in conditional clustering}
\label{subsubsec:conditional_4}
Landscape 4 of the conditional clustering has a distribution similar to
that of the corresponding landscape 3.
The variability is higher than in the case of landscape 3 of this clustering.

\vspace*{-1pt}
\subsubsection{Distribution in clustering $C^{(412)}$}
\label{subsubsec:412_4}
This landscape is present mainly along the north-western and the
mid-eastern region of the Western Ghats,
with variability higher than that of landscape 3 of $K$-means and the
conditional clustering.
The vegetation is mainly dry deciduous in the north-west and wet
evergreen in the mid-east.

\subsubsection{Distribution in clustering $C^{(759)}$}
\label{subsubsec:759_4}
The distribution of landscape 4 of $C^{(759)}$ is very similar to that
of landscape 4 of $C^{(412)}$,
but the variability is higher.

\subsection{Landscape type-5}
\label{subsec:land_5}
\subsubsection{Distribution in $K$-means clustering}
\label{subsubsec:kmeans_5}
As in the case of landscape 5 of NG, here also landscape 5 is
distributed along the entire length
of the Western Ghats, but more toward the western side, rather than the
eastern side as found by NG
in their landscape 5. A number of individual landscape elements are
indicated by the mean standard deviation.

\subsubsection{Distribution in conditional clustering}
\label{subsubsec:conditional_5}
Landscape 5 associated with the conditional clustering is distributed
along the entire Western Ghats, and has
smaller variability than landscape 5 of the $K$-means clustering.

\subsubsection{Distribution in clustering $C^{(412)}$}
\label{subsubsec:412_5}
For $C^{(412)}$ landscape 5 is present mainly in the eastern parts and
in the southern foothills.
The mean standard deviation is even smaller than landscape 5 of the
conditional clustering.
The natural vegetation is wet evergreen and moist deciduous forests.

\subsubsection{Distribution in clustering $C^{(759)}$}
\label{subsubsec:759_5}
The distribution of landscape 5 of $C^{(759)}$ resembles that of
landscape 5 of $C^{(412)}$, although
the distribution of the former is less prominent in the eastern side
and the southern foothills.
The mean standard deviation is not that significant, although it is
higher than in
landscape 5 of $C^{(412)}$.

\subsection{Landscape type-6}
\label{subsec:land_6}
\subsubsection{Distribution in $K$-means clustering}
\label{subsubsec:kmeans_6}
The distribution of landscape 6 of the $K$-means clustering is over the
entire Western Ghats, similar to the
distribution of landscape 6 of NG, but toward the south it is
distributed more in the west, rather than
in the east, as in NG. In the north, the distribution is more toward
the east, rather than toward the west coast,
as in NG. The mean standard deviation being 1.48 is not that significant.

\subsubsection{Distribution in conditional clustering}
\label{subsubsec:conditional_6}
Landscape 6 of the conditional clustering is distributed along the
entire length of the Western Ghats,
with higher mean standard deviation compared to landscape 6 of the
$K$-means clustering.

\subsubsection{Distribution in clustering $C^{(412)}$}
\label{subsubsec:412_6}
Here the distribution is again over the entire Ghats, but with larger
mean standard deviation
compared to landscape 6 of the conditional clustering.

\subsubsection{Distribution in clustering $C^{(759)}$}
\label{subsubsec:759_6}
The distribution of landscape~6 of $C^{(759)}$ is mainly in the
northern, north-western and mid-western region
of the Western Ghats, with significantly high mean standard deviation,
suggesting a large number of individual
landscape elements. The natural vegetation is dry deciduous and evergreen.

\subsection{Landscape type-7}
\label{subsec:land_7}
\subsubsection{Distribution in $K$-means clustering}
\label{subsubsec:kmeans_7}
Very closely resembling landscape~7 of NG, landscape type 7 of the
$K$-means clustering is distributed
both to the east and west of the entire Western Ghats. Here the natural
vegetation is of the wet evergreen type,
extending to moist deciduous in the southern part of the Western Ghats.
It is this landscape within which, according to NG,
most evergreen forests of the Western Ghats fall.
The natural vegetation has been replaced to a large extent by woodland
to savanna-woodland, tree savanna to grass savanna,
thickets and scattered shrubs.
As for the crops, millets, cotton and rice are grown in the north
while millets and oilseeds are grown in the south. Arecanut, coconut,
coffee, etc. are also grown
in this landscape. The mean standard deviation being 1.68 does not
indicate a large number of ecosystems.

\subsubsection{Distribution in conditional clustering}
\label{subsubsec:conditional_7}
Landscape 7 of the conditional clustering is again distributed all over
Western Ghats. The mean standard deviation
is somewhat large, suggesting quite a few individual landscape elements.

\subsubsection{Distribution in clustering $C^{(412)}$}
\label{subsubsec:412_7}
The distribution of landscape 7 of $C^{(412)}$ resembles that of
landscape 7 of the conditional clustering.
The mean standard deviations are also similar.

\subsubsection{Distribution in clustering $C^{(759)}$}
\label{subsubsec:759_7}
Landscape type 7 of $C^{(759)}$ resembles landscape type 7 of the
conditional clustering and $C^{(412)}$, but
it is distributed more prominently toward the north-east and the
southern parts of the Ghats. The natural vegetation
is mainly dry deciduous and evergreen, extending to moist deciduous in
the south.
The mean standard deviation being small does not indicate too many ecosystems.

\subsection{Landscape type-8}
\label{subsec:land_8}
\subsubsection{Distribution in $K$-means clustering}
\label{subsubsec:kmeans_8}
Landscape 8 with respect to the $K$-means clustering is mainly present
in the western part of the northern regions
and both eastern and western parts of the middle and southern regions.
This is unlike the distribution of landscape 8
of NG, which is present mostly in the western part and absent in the
north. Rather, the distribution of
landscape 8 of the $K$-means clustering resembles landscape 7 of the
$K$-means clustering and that of NG.
The mean standard deviation is, however,
higher in this case.

\vspace*{-2pt}
\subsubsection{Distribution in conditional clustering}
\label{subsubsec:conditional_8}
The distribution of landscape 8 of the conditional clustering closely
resembles the distributions of the previous
landscapes of the same clustering.
The mean standard deviation does not indicate too many ecosystems.

\vspace*{-2pt}
\subsubsection{Distribution in clustering $C^{(412)}$}
\label{subsubsec:412_8}
Landscape type 8 of $C^{(412)}$ is present most in the northern and
north-western regions of the Western Ghats.
The vegetation is mostly dry deciduous. The variability is significant,
indicating many ecosystems.

\vspace*{-2pt}
\subsubsection{Distribution in clustering $C^{(759)}$}
\label{subsubsec:759_8}
Landscape type 8 of $C^{(759)}$ is distributed mainly along the
northern, north-western and mid-western
regions of the Ghats. The vegetation is mainly dry deciduous, extending
to evergreen. The variability is high,
suggesting many ecosystems.

\subsection{Landscape type-9}
\label{subsec:land_9}
\vspace*{-2pt}
\subsubsection{Distribution in $K$-means clustering}
\label{subsubsec:kmeans_9}
Agreeing very closely with NG, landscape type 9 of $K$-means clustering
is nearly absent in the northern
stretches and is present in the central and southern parts in patches
toward the west. Here the natural
vegetation is evergreen and semi-evergreen. Disturbed semi-evergreen
forests along with moist deciduous forests,
woodlands and savanna-woodlands are also present. Crops like rice,
tapioca, coconut, millets and oilseeds are grown.
Relatively high mean standard deviation suggests a mixture of several
ecosystems.

\vspace*{-2pt}
\subsubsection{Distribution in conditional clustering}
\label{subsubsec:conditional_9}
The distribution of landscape 9 of the conditional clustering is all
over the Ghats, but in the mid-western
regions the distribution is more prominent. The vegetation in this
region is evergreen. The variability suggests
several individual landscape elements.

\vspace*{-2pt}
\subsubsection{Distribution in clustering $C^{(412)}$}
\label{subsubsec:412_9}
Landscape 9 of $C^{(412)}$ is distributed all over the Ghats but is
present more prominently
toward the eastern parts. Not many individual landscape types are
indicated by the mean standard deviation.

\vspace*{-2pt}
\subsubsection{Distribution in clustering $C^{(759)}$}
\label{subsubsec:759_9}
Landscape 9 of $C^{(759)}$ is distributed all over the Ghats but is
present more prominently
toward the western parts. Quite a few individual landscape types are
indicated by the mean standard deviation.

\subsection{Landscape type-10}
\label{subsec:land_10}
\vspace*{-3pt}
\subsubsection{Distribution in $K$-means clustering}
\label{subsubsec:kmeans_10}
Landscape type 10 is present in a few patches toward the northern and
in the central parts of the Ghats.
The vegetation is evergreen. The low mean standard deviation does not
suggest much heterogeneity.

\vspace*{-3pt}
\subsubsection{Distribution in conditional clustering}
\label{subsubsec:conditional_10}
In the conditional clustering, landscape type 10 is found mainly in the
north-western, mid-western and
in the southern parts of Western Ghats. The vegetation ranges between
dry deciduous, evergreen and moist deciduous.
Here also relative homogeneity is indicated by the low mean standard deviation.

\vspace*{-3pt}
\subsubsection{Distribution in clustering $C^{(412)}$}
\label{subsubsec:412_10}
In $C^{(412)}$ landscape type 10 is present mainly in the western part
along the entire length of the Ghats.
The natural vegetation is deciduous as well as evergreen. Among crops,
rice, tapioca and coconut are planted.
The variability suggests nonnegligible heterogeneity.

\vspace*{-3pt}
\subsubsection{Distribution in clustering $C^{(759)}$}
\label{subsubsec:759_10}
Landscape type 10 with respect to $C^{(759)}$ is mostly present in the
mid-eastern regions and the southern part
of the Western Ghats. The natural vegetation is wet evergreen and moist
deciduous. Not much heterogeneity is indicated
by the mean standard deviation.

\subsection{Landscape type-11}
\label{subsec:land_11}
\vspace*{-3pt}
\subsubsection{Distribution in $K$-means clustering}
\label{subsubsec:kmeans_11}
$\!$In contrast with landscape type~11 of NG, which is present in a single
patch, here it is present in the
northern stretches, and in the eastern stretches of the central and the
southern parts of Western Ghats.
The vegetation varies from dry deciduous to moist deciduous forests,
with wet evergreen forests in the
mid-eastern parts. A fair amount of heterogeneity is indicated by the
mean standard deviation.

\vspace*{-3pt}
\subsubsection{Distribution in conditional clustering}
\label{subsubsec:conditional_11}
Landscape type 11 associated with the conditional clustering is present
mainly in the mid-eastern and the southern parts of the Ghats.
The vegetation here is mostly wet evergreen and moist deciduous. A fair
amount of homogeneity is indicated by the mean
standard deviation.

\vspace*{-3pt}
\subsubsection{Distribution in clustering $C^{(759)}$}
\label{subsubsec:759_11}
Landscape 11 of $C^{(759)}$ is present all over Western Ghats but is
more prominent toward the mid-eastern
and the southern parts, as in the case of landscape 11 of the
conditional clustering.
A fair amount of individual landscape types are indicated by the variability.

\subsection{Landscape type-12 of $C^{(759)}$}
\label{subsec:land_12}
Landscape type 12 of $C^{(759)}$, in spite of its presence all over the
Ghats, is more prominent
in the mid-western and the south-western stretches. The natural
vegetation is mainly evergreen
and moist deciduous. Again, a fair amount of individual landscape types
are suggested by the variability.

\subsection{Landscape type-13 of $C^{(759)}$}
\label{subsec:land_13}
This landscape is present mainly in the central and the southern parts
of the Ghats, with mostly evergreen
and moist deciduous vegetation. A very low mean standard deviation
indicates homogeneity.

\subsection{Landscape type-14 of $C^{(759)}$}
\label{subsec:land_14}
This landscape is present all over Western Ghats, but mainly along the
western stretches and in the southern
foothills. The vegetation
is dry deciduous in the north, evergreen in the center and moist
deciduous in the south.
The mean standard deviation indicates some amount of heterogeneity.

\section{Conclusion}
\label{sec:conclusion}

We have highlighted the importance of acknowledging clustering
uncertainty, and have introduced methodologies for
summarizing posterior distributions of clusterings. We have
demonstrated how central clusterings can be obtained
from posterior samples drawn using MCMC methodologies. In the heart of
our proposed methods for summarizing posterior distributions of
clusterings is
a clustering metric introduced to compare any two clusterings. Although
computation of the exact distance between two clusterings can be expensive,
we have introduced a computationally cheap, and perhaps, more
importantly, accurate, approximation to the exact metric.
We remark that although we have confined our attention to the modes of
the posterior distribution of clusterings in this paper,
it is also possible to obtain the median of the posterior distribution
of clusterings.
For instance, the median $C^{(\mathrm{med})}$ may be defined as
$C^{(\mathrm{med})}=\arg\min_C\sum_{i=1}^N\hat d(C^{(i)},C)$. Also,
considering any ``reference clustering'' $C^{(0)}$,
acting as the origin, a partial ordering ``$\preceq$'' with respect to
the origin can be defined on the set of
the clusterings obtained from MCMC sampling as $C^{(i)}\preceq C^{(j)}$
if and only if
$\hat d(C^{(0)},C^{(i)})\leq\hat d(C^{(0)},C^{(j)})$, for any~$i,j$.
Using this partial ordering, any
quantile with respect to the origin can be calculated.

Analysis of the Western Ghats data
based on our proposed methodologies revealed broad similarities with
the results obtained by NG, which includes
the number of clusters obtained by them is the one which gets the
highest posterior probability corresponding to our Bayesian model. However,
we have also pointed out that the clustering obtained by NG is unlikely
to fall within our unconditional 95\% credible or HPD regions,
although it is likely to fall within our conditional 95\% credible or
HPD regions, conditioned on the number of clusters,
fixed in their deterministic algorithm. Such a drawback, as we have shown,
is due to the failure of the deterministic algorithm to take account of
the uncertainty involved with
the number of clusters. The detailed results of our application to the
biodiversity hotspot Western Ghats reveal dissimilarities of the
landscape types obtained by our
clustering methodology with that obtained by a proxy to NG's clustering
algorithm. As to be expected, some similarity is exhibited between the
landscape types
obtained by these methods, when we conditioned on the same number of
clusters fixed by NG. These new and interesting facts indicate that
ecologists may need to update their
methodologies for studying biodiversity. The methodologies we proposed
in this paper are not expected to be immediately accessible to
ecologists because
of the technical gap between ecological and statistical communities,
but collaborative efforts may yield fruit in the future, benefitting
both communities.

\vspace*{5pt}
\section*{Acknowledgments}

We are grateful to Professor Madhav Gadgil and Dr. Harini Nagendra for
providing the Western Ghats data set to
Professors Jayanta Ghosh and Tapas Samanta and to the latter for
sharing this data with us and
suggesting the clustering problem, providing valuable insights and
active involvement
in a previous version of this paper. The authors very gratefully
acknowledge the help provided by the Editor, an
Associate Editor and a referee, whose very detailed and thorough
comments led to a greatly improved presentation of this paper.

\vspace*{5pt}
\begin{supplement}[id=suppA]
\stitle{Supplement to ``On Bayesian ``central clustering'':
Application to landscape classification of Western Ghats}
\slink[doi]{10.1214/11-AOAS454SUPP} 
\slink[url]{http://lib.stat.cmu.edu/aoas/454/supplement.pdf}
\sdatatype{.pdf}
\sdescription{Sections S-1 and S-2 contain, respectively, the full
conditional distributions of the
random variables with respect
to the nonmarginalized and marginalized version of SB's model. That the
$K$-means clustering
algorithm is a special case
of the clustering method based on SB's model is shown in Section S-3.
Properties of the approximate distance
measure $\hat d$ are explored in Section S-4. Section S-5 contains
reports of our investigation on whether
or not the spatial structure of the superpixels should be incorporated
in our model. Detailed analysis of
sensitivity of the results with respect to changes in the values of the
hyperparameters of our model
is provided in Section S-6. Thorough explanation of the computational
superiority of SB's model
over that associated with efficient implementation of EW's model is
presented in Section S-7.
Finally, a new method for MCMC convergence diagnostics in clustering
models is proposed
in Section S-8, which we apply in our situation for studying convergence
of our Markov chain.}
\end{supplement}

%

\printaddresses

\end{document}